\documentclass[a4paper,fleqn,usenatbib,useAMS]{mnras}
\usepackage{graphicx}
\usepackage{natbib}
\usepackage{color}
\usepackage{amsmath}
\usepackage{amssymb}
\usepackage{multicol}   
\usepackage{bm}		
\usepackage{pdflscape}	

\newcommand{\ie}{i.\,e.}

\newcommand{\eg}{e.\,g.}

\newcommand{\teff}{$T_{{\rm eff}}$}
\newcommand{\kms}{\mbox{km\,s$^{-1}$}}
\newcommand{\ms}{\mbox{m s$^{-1}$}}

\newcommand{\fwhm} {{\sc fwhm}}

\newcommand{\snr} {\mbox{SNR}}

\newcommand{\gfeh}{\mbox{$[{\rm Fe}/{\rm H}]$}}

\newcommand{\logg}{\mbox{log~{\it g}}}

\usepackage[T1]{fontenc}
\usepackage{ae,aecompl}

\title[The M~4 Core Project with \textit{HST} - IV]{The M~4 Core Project with
  \textit{HST} - IV. Internal Kinematics from Accurate Radial
  Velocities of 2771 Cluster Members\thanks{Based on
 observations  collected at  ESO  Paranal  Observatory within  the
observing program 71.D-0205, 77.D-0182 and 383.D-0802}}

\author[L.\ Malavolta]{L.\ Malavolta$^{1,2}$\thanks{Corresponding authors: e-mail:
luca.malavolta@unipd.it (LM)},
G.\ Piotto$^{1,2}$,
L.\ R.\ Bedin$^2$, 
C.\ Sneden$^3$,\newauthor  
V.\ Nascimbeni$^2$,
V.\ Sommariva$^{4,5}$\\
$^{1}$Dipartimento di Fisica e Astronomia ``Galileo Galilei'', Universit\`a di Padova, vicolo dell'Osservatorio 3, Padova IT-35122 \\
$^{2}$INAF - Osservatorio Astronomico di Padova, vicolo
dell'Osservatorio 5, Padova, IT-35122 \\
$^{3}$Department of Astronomy and McDonald Observatory,
The University of Texas, Austin, TX 78712, USA \\
$^{4}$Dipartimento di Fisica e Astronomia - DIFA,  Universit\`a di Bologna,
viale Berti Pichat 6\/2, Bologna, IT-40127 \\
$^{5}$INAF - Osservatorio Astronomico di Roma, via Frascati 33, Monte
Porzio Catone,  IT-00040 \\
}

\begin{document}

\pagerange{\pageref{firstpage}--\pageref{lastpage}} \pubyear{2015}

\maketitle
\label{firstpage}
\begin{abstract}

We present a detailed study of the internal kinematics of the Galactic 
Globular Cluster M\,4 (NGC~6121), by deriving the radial velocities from 
7250 spectra for 2771 stars distributed from the upper part of the Red 
Giant Branch down to the Main Sequence. 
We describe new approaches to determine the wavelength 
solution from day-time calibrations and to determine the radial velocity 
drifts that can occur between calibration and science observations when 
observing with the GIRAFFE spectrograph at VLT.

Two techniques to determine the radial velocity are
  compared, after a qualitative description of their advantages with
  respect to other commonly used algorithm, and a new approach to remove the sky contribution from the
  spectra obtained  with fibre-fed spectrograph and further improve
  the radial velocity  precision is presented. 
The average radial velocity of the cluster is $\langle
  v \rangle = 71.08 \pm 0.08$~\kms\ with an average dispersion of
  $\mu_{v_c} = 3.97$~\kms. Using the same dataset and the same statistical 
approach of previous analyses, 20 additional binary candidates are
found, for a total of 87 candidates. 
A new determination of the internal radial velocity dispersion as a function of
cluster distance is presented, resulting in a dispersion of $4.5$~\kms\ within
2$\arcmin$ from the center of cluster and steadily decreasing 
outward. We statistically confirm the small amplitude of the cluster rotation, as suggested in the past by several authors.
This new analysis represents a significant improvement with
respect to previous results in literature and provides a fundamental
observational input for the modeling of the cluster dynamics.

\end{abstract} 


\section{Introduction}\label{sec:intro}

M\,4 (NGC~6121) is one of the most studied Galactic 
Globular Clusters (GGC). 
Its proximity, brightness and the moderate concentration make possible 
the study of its inner regions in great detail. 
This cluster has been the object of several intensive photometric
campaigns, both from space (\eg\
\citealt[Paper~I]{Bedin:2013cd}, \citealt{Piotto:2015uv}) and from 
the ground (\eg\ \citealt{Kaluzny:2013cd}, \citealt{Libralato:2014m4}).
Red giant members of this cluster are accessible to high-resolution
spectroscopy with even medium-size telescopes, resulting in a wealth
of abundance studies (\eg\ \citealt{Brown:1992a}, \citealt{Ivans:1999hf},
\citealt{Smith:2005a}, \citealt{Yong:2008a}, \citealt{Marino:2008du},
\citealt{Dorazi:2013a}), 
which have also provided evidences of multiple populations. Only
recently the presence of multiple populations has been photometrically
observed in the Red-Giant Branch (RGB) \citep{Monelli:2013ab} and
in the Main Sequence (MS) (\citealt[Paper~II]{Milone:2014ab},  \citealt{Nardiello2015:m4})
This cluster also represents a natural benchmark for many theoretical
studies, such as numerical simulations of the internal dynamics
\citep{Heggie:2008c5,Heggie:2014m4} and the interactions with the Galaxy 
(\eg\, \citealt{Dinescu:1999cd}).

Despite being so well-studied, the internal dynamics 
of this cluster are still poorly known.  
The kinematics of other GGC's have been explored in more detail.
One purpose of such studies has been to determine the presence of
a central intermediate-mass black hole in a cluster center, using 
integral-field spectroscopy to measure the radial velocity dispersion in 
the center of the cluster (e.g. within 10$\arcsec$; see for
example \citealt{Feldmeier:2013cd}, \citealt{Lutzgendorf:2013cd}),
Theories of gravity have also been tested by measuring radial
velocities of individual stars in the outer regions of a cluster
(farther than 1$\arcmin$ from the center of the cluster, see for
example \citealt{Baumgardt:2009cd}, \citealt{Ibata:2011cd}).
Both techniques have their disadvantages when determining the kinematical
status of a cluster: integral field spectroscopy requires a
deconvolution of the observed spectra along the line of sight, while
results from individual stars may be hampered by the low statistics,
especially near the center of far clusters where it may hard to
observe individual stars.  

M\,4 represents an excellent target for 
kinematic analysis, thanks to its proximity and the many detailed 
studies of its properties. 
Still, the latest determination of the internal radial velocity
dispersion as a function of distance dates back to the
radial velocity study of Peterson, Rees \& Cudworth (1995, hereafter  P95) 
\nocite{Peterson:1995iy}, which sampled 
182 members with radial velocity precision 1~\kms.
Despite the many recent spectroscopic campaigns on M\,4, more precise 
and accurate wavelength calibrations are required for radial velocity 
determination than for chemical analyses.
Spectra obtained for abundance studies may be affected by
systematic shifts in radial velocity among different observations, or simply
lack the required spectral resolution for a precise determination of
velocities.
These aspects are usually not important for chemical analyses, but
they severely impact the use of spectroscopic data from different
instruments for detailed kinematics analyses.

During last decade our group has collected thousands 
of spectra using the GIRAFFE spectrograph at VLT to estimate the geometrical
distance, by comparing the radial velocity dispersion with the proper
motion dispersion, and dynamical state of several GGC.
Analyses of these spectra have been hampered by systematics in data 
calibration and radial velocity measurements; 
see for example \cite{Milone:2006cc} and 
Sommariva et~al. (2009; hereafter S09)\nocite{Sommariva:2009cz}.

This project is the spectroscopic counter-part of an
extensive astrometric and photometric investigation
of M\,4, involving about
700 WFC3/UVIS and 120 ACS/WFC \textit{HST} images (GO-12911, PI Bedin), aimed at searching for and
characterizing the binary population in the core as well as in
the outskirts of the cluster (\citealt[Paper~I]{Bedin:2013cd} for a careful description of the
\textit{HST} large program GO-12911).

In this paper we have focused our attention on significantly increasing 
wavelength calibration accuracies and radial velocity determinations 
of very low \snr\ stellar spectra.
After introducing the dataset in \S\ref{sec:data-description},
we devote \S\ref{sec:giraffe-wavel-calibr} to a detailed description of 
the improvements in wavelength calibration.
Our methodology to properly account for the geometrical distortions in 
a CCD chip when determining the drift in radial velocity that may occur 
between day-time calibration and science observations is
described in \S\ref{sec:simcal-calibration}. 
In \S\ref{sec:radi-veloc-determ} three well-known techniques are
compared in order to establish which one is the most suitable to the
characteristics of the instrument.

Finally, in \S\ref{sec:results-discussion} we discuss
the application of these improved data reduction techniques to a dataset 
comprising 7250 spectra obtained with GIRAFFE for 2771 stars of the 
GGC M\,4 in the wavelength region $5145 - 5360 \,$\AA. 
Only about 300 stars in our sample reside on the RGB.
The vast majority of the stars are members of the Sub-Giant Branch (SGB) and 
MS stars, \ie\ relatively faint targets with low \snr\
spectra.  
Careful consideration of the continuum placement and the sky contribution 
for each stellar spectrum is required.
These aspects have already been discussed in 
Malavolta et~al. (2014; hereafter LM14)\nocite{Malavolta:2014hu}.
While in our previous work we focused on the determination of 
atmospheric parameters over a wide range of evolutionary states and
spectra of different quality, in this paper we aim to increase the
precision of the radial velocities available in the literature
 and provide new insights into its internal kinematics.

\section{The Spectroscopic Data Set}\label{sec:data-description}

The spectra for our project were originally used by
S09 to study the internal velocity
dispersion of M\,4 and to search for spectroscopic binaries.
As described in LM14, the instrumental configuration was
the GIRAFFE medium-high resolution spectrograph fed by the VLT Fibre 
Large Array Multi Element Spectrograph (FLAMES; \citealt{Pasquini:2000tk}) 
in MEDUSA multi-fibre mode.
This configuration produced single-order spectra for 
132 objects (target stars and sky) in each integration.
Using the GIRAFFE HR9 setup yielded spectra with dispersion of 
0.05~\AA/pixel  and measured 4-pixel resolving power 
R~$\equiv$~$\lambda/\Delta\lambda$~=~25,800 in the 
wavelength range $5143$\AA\ $< \lambda < 5356$\AA. 
The HR9 wavelength region is particularly useful for radial velocity 
studies because it contains a large number of iron lines and very 
strong absorption lines, notably the Mg-b triplet. 

The M\,4 target stars were selected by S09 from an astrometric and
photometric catalog based on Wide Field Imager (WFI) data from the
ESO/MPIA 2.2m telescope \cite{Anderson:2006a}.
Each fibre's radius on the sky is 
0.6$\arcsec$, so each chosen star was constrained 
to have no bright neighbors (defined as $V_{neighbor} < V_{target} + 2.5$)
within an angular distance of 1.2$\arcsec$.

The spatial distribution and photometric properties of our target stars are
the same as illustrated in Fig.~1 and 2 of LM14.
Our target stars essentially sample all of M\,4 spectroscopically up
to twice the half-mass radius.
The limiting magnitude of the targets, $V \lesssim 17.5$, was
chosen to ensure that a single M\,4 integration had signal-to-noise 
$\textrm{\snr}>10$ for each star, which
was considered by S09 to be the
minimum threshold to determine radial velocities (RVs) with a precision of a
few hundred  m~s$^{-1}$, on the basis of previous experience with the instrument.

A total of 2771 stars covering CMD positions 
from the upper red-giant branch to about one magnitude fainter than the 
main-sequence turnoff (TO) luminosity were observed
between 2003 and 2009, including 306 new spectra obtained in
2009 and targeting MS stars already observed in the previous epochs. 
Determination of the M\,4 velocity dispersion and binary star 
fraction were the prime motivators for obtaining these data.
Therefore nearly all stars were observed at least twice, and three 
or more spectra were obtained for nearly 40\% of the sample.
A total of 7250 individual spectra were available for our study,
as we summarize in Table~\ref{tab-nobs}.

\begin{table}
\center
\caption{Spectroscopic Observation Statistics} 
\label{tab-nobs} 
\begin{tabular}{ccc} 
\hline
No. Observations & No. Stars & No. Total \\
\hline\hline
  1  &        83     &      83     \\
  2  &      1722     &    3444     \\
  3  &       501     &    1503     \\
  4  &       321     &    1284     \\
  5  &        72     &     360     \\
  6  &        20     &     120     \\
  7  &         1     &       7     \\
  8  &        27     &     216     \\
  9  &        10     &      90     \\
 10  &        12     &     120     \\
 11  &         1     &      11     \\
 12  &         1     &      12     \\
     &               &             \\
Total &     2771     &    7250     \\
\hline
\end{tabular}
\end{table}

Default basic reductions (bias and dark correction, flat fielding) have been performed with the ESO pipeline. 
Spectral extraction has been performed using the optimal spectral extraction algorithm \citep{Horne:1986bg} and the empirical point-spread function concept from photometry \citep{Anderson:2000bt,Anderson:2006a} adapted to spectroscopy (LM14). The extracted spectra are fully consistent with the ones delivered by the ESO pipeline using either the \textrm{HORNE} option or the standard \textrm{SUM}, as a consequence of the fact that scattered light and fiber cross-talk are negligible in this specific setting of the MEDUSA mode.
Wavelength calibration is performed by taking a hollow cathode
Thorium-Argon lamp (hereafter simply Th-Ar lamp) spectrum with each of
the fibres in the MEDUSA plate with the 
selected instrument setup. 
These standard calibration frames are usually taken during the day, 
several hours before observations.
During a science exposure five fibres called 
\textit{simultaneous calibration fibres} 
(SimCals) can be dedicated to take additional 
Th-Ar spectra, in order to correct for changes in the wavelength 
dispersion solution caused by variations in air 
pressure and temperature occurred inside the spectrograph
between the day-time calibration and night observations.
S09 used the ESO Giraffe standard reduction pipeline 
\citep{Blecha:2000ud} and the Ancillary Data Analysis Software
\citep{Royer:2002hf} to reduce the M\,4 data from raw CCD exposures to
wavelength-calibrated 1D spectra, apply the drift correction
in the wavelength scale and determine the radial velocities of the
observed stars. 

A systematic offset of 150~\ms\ in the radial velocity zero point was
observed by in the five simultaneous calibration fibres between 2003 and
2006 data (see \S 4.1 of S09 for a detailed discussion).
Changes in environmental conditions inside the spectrograph (\ie\ air 
pressure and temperature), as well as switching to different
  instruments modes, could cause an overall shift in wavelength (and hence RV) in the
interval of time between observations and 
calibration exposures and between different nights.
These should be automatically taken into account by the ESO
  pipeline using the simultaneous calibration fibres.
After removing this source of systematic error, the only expected
variation in RVs of calibration spectra taken in different epochs should have a random 
distribution, given by the precision in RV of the spectrograph and independent with time.

The dispersion inside a single epoch is several times smaller then 
the offset (Fig.~6 of S09),
and this seems to 
imply a problem in the original wavelength dispersion solution. 
Therefore we decided to revisit the basic reductions of the raw data
sets, to see if the velocity precision could be significantly improved.
In the next sections our different approach to wavelength calibration 
and drift correction is described. 
To mark the difference with the ESO pipeline, we will refer to our 
approach as \textit{Direct Calibration}.

\section{GIRAFFE Wavelength Calibration}\label{sec:giraffe-wavel-calibr}

ESO provides a pipeline for GIRAFFE as a part of the 
\textit{VLT Data Flow System} (DFS). 
A detailed description of the pipeline can be found at the ESO 
website\footnote{\url{http://www.eso.org/sci/software/pipelines/}}; 
here we provide a brief
description of the wavelength calibration subroutine to highlight the
differences with our \textit{Direct Calibration} technique.

\subsection{ESO Pipeline calibration}\label{sec:eso-pipel-calibr}

Calibration frames used for dispersion solutions are obtained by
illuminating the entrance of the fibres with a hollow cathode
Th-Ar lamp
using the same setup as for the stellar spectra. 
A complete GIRAFFE dispersion solution consists of an optical model of
the fibre spectra onto the CCD detector, a
polynomial fit of the optical model residuals, and the correction of
individual fibre offsets. 

The solution is determined iteratively, \ie\ at each iteration the previous 
model is updated until the measured emission lines have a negligible radial
velocity with respect to the tabulated values. 
A previous dispersion solution is useful for rapid computations, but, if
necessary, a new dispersion solution can be computed from scratch using the 
average optical model that 
accompanies the pipeline software.

In detail, for each wavelength calibration procedure, a set of 
unsaturated Th-Ar emission lines is selected according to the instrumental 
setup of the images.  
The position of these lines on the CCD is determined using an 
analytical optical model. 
The residuals between these positions and the measured ones are
then modeled with a 2D Chebyshev polynomial. 
The global wavelength solution is then the sum of the
  optical model and of the polynomial residuals. See
  \cite{Royer:2002hf} for a detailed description of the ESO pipeline,
  and \cite{Bristow2010:sp} on the use physical
models for instrument calibration.

\subsection{Direct Calibration}\label{sec:direct-calibration}

 A more classical, direct approach has been followed in our work. 
For each individual Th-Ar spectrum, a selected list of
emission lines with known wavelengths are measured and then a 
piecewise polynomial (represented as Basis spline, \citealt{Boor:1977:PCB})
is fitted to determine the wavelength dispersion solution.

The algorithm presented here is designed to automatically perform those 
steps that in other tools such \textrm{IRAF}\footnote{
IRAF is distributed by the National Optical Astronomy 
Observatory, which is operated by the Association of Universities for 
Research in Astronomy (AURA) under cooperative agreement with the 
National Science Foundation.}
require human interaction.
Although our reference wavelength range and resolution are the ones 
defined by the HR9 setting of GIRAFFE, we want to keep our algorithm 
as general as possible for rapid adaptation to different wavelength 
GIRAFFE ranges or for spectra obtained with different instruments.

\subsubsection{Line lists of Thorium and Argon lines}\label{sec:line-list-thorium}
Several published Th-Ar line lists are available in which lines have been 
selected according to their stability and strength from high resolution
spectra; see for example \cite{Murphy:2007hn,Lovis:2007ac}. 
However, when dealing with medium resolution spectroscopy,
blending with close or faint lines becomes important. 
The centers of lines of interest can be shifted several hundredths
of \AA ngstroms, thus leading to a less precise determination of the 
wavelength dispersion even with excellent data and a very accurate line list.

To create a line list that takes into account line blending, we
proceeded in the following way. 
We have selected the Th-Ar reference spectrum obtained with ESO's 3.6m telescope Coude 
Echelle Spectrometer (CES, \citealt{Enard:1982wf}) as our reference
among several Th-Ar spectrum available in the literature after
checking the similarity with GIRAFFE Th-Ar spectra (emission features
may change depending on the manufacturer of the lamp).
This spectrum has been obtained with a resolving power of $R=200,000$, 
which is almost nine times the resolving power of GIRAFFE in the HR9 
setting. 
We degraded this spectrum to match the resolution of our spectra; we will 
refer to this smoothed spectrum as the \textit{reference} one. 
All the calibrations will be relative to
the reference system given by this spectrum.

\begin{figure}
\includegraphics[width=\columnwidth]{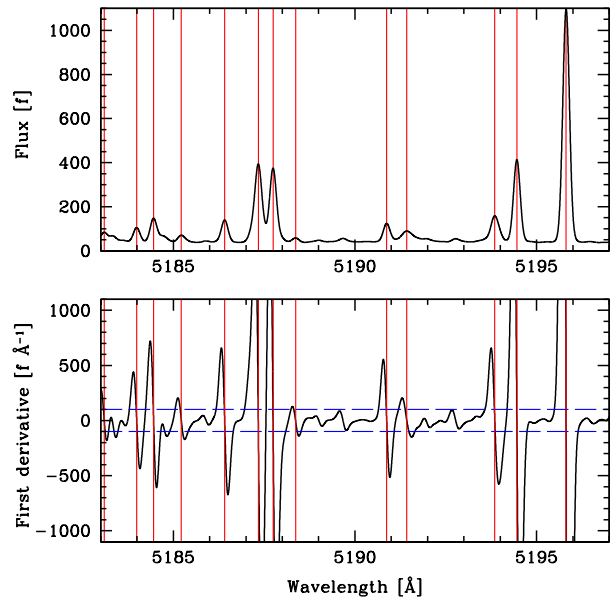}
\caption{An example of the line identification on the Thorium-Argon 
  reference spectrum, with the smoothed spectrum 
  in the upper panel and its first derivative in the lower one. 
  Very weak lines have been excluded in our analysis if the inflection points of their
  first derivative are within the two blue dashed lines. Selected
  lines are denoted by a red vertical line.}
\label{fig:thar}
\end{figure}

Lines have been automatically identified using the first derivative of
the reference spectrum. 
Fig.~\ref{fig:thar} shows an example of the line identification 
process, with the smoothed reference spectrum in the upper panel (in
arbitrary units $f$) and its first derivative in the lower one (in arbitrary flux 
units per \AA ngstrom $f $\AA$^{-1}$). 
To separate strong lines from small features and weak
  emission lines, we have selected only lines with null derivative at
  the expected central wavelength  
and inflection points on the side that have a minimum absolute value
of 100 $f$\AA$^{-1}$.

In our wavelength range 178 lines have been identified. The positions
of these lines have been measured on the reference spectrum, and their
values define our rest-frame system of reference for the determination
of the wavelength dispersion solution of GIRAFFE calibration spectra.

\subsection{First approximation of the wavelength dispersion
  solution}\label{sec:first-appr-wavel}

A first approximation to the dispersion solution is needed to
allow the algorithm to determine the solution. 
 
We start by selecting the ten lines with the largest fluxes
from the reference Th-Ar spectrum in the wavelength range of interest. 
The number of lines has been chosen in a way that the included lines form 
a subset with mean flux that is at least a few times the flux of the
brightest lines not included in the subset. 
This ensures that the selection in the observed spectra will not be affected 
by differences in the line flux ratios that may be present when comparing 
spectra obtained with lamps made by different manufacturers. 
A very rough pixel position for each line must be provided 
also: a single value is good enough for all the fibres, 
despite their shift in the cross-dispersion direction. 
The wavelength values are used as initial guesses for a Gaussian fit, so a 
precision of one half of the full-width-at-half-maximum (\fwhm) is good enough. 
This selection can be easily performed by visual
  inspection directly at the
reference spectrum, and must be done only once for a given wavelength
range or instrument setup. 
This is the only human interaction required by the algorithm. 

A Gaussian fit 
on these pre-selected lines 
is performed on both the pixel space for the 
observed
spectrum
and in the wavelength space for the reference one. 
The values are sorted in increasing order, so that 
the values in the pixel space have now an associated wavelength. 
The only precaution to be taken at this step is to check that no strong 
lines are present close to the border of the wavelength range 
(\eg\ $10$ \AA\ from each limit for our instrument setting), otherwise 
an error in the association of lines 
with pixels can occur.
In fact, as a result of the optical design of GIRAFFE, the 
wavelength range of the spectra varies slowly across the CCD, with a maximum
shift of about $10$ \AA\ for the fibres close to the center of the sensor respect to the
first one.

Before proceeding to the first approximate determination of a wavelength 
solution, we need an additional step to ensure that cosmic rays or 
uncorrected hot pixels have not been identified as spectral lines. 
By inspection of our data frames, we noticed that real spectral
lines have a total shift of around 100 pixels in the cross-dispersion
direction (\ie\ across the fibres). 
The maximum shift with respect to the first fibre around the center of the
frame is 50 pixels, so we set this value as a threshold for checking
proposed line identifications in the various fibres. 
This steady shift in line positions in different fibres is a
consequence of the optical design of the instrument and it varies
smoothly across the sensor, so for each wavelength value we perform a
low-degree polynomial interpolation
of the pixel values versus their fibre number. 
The fit is performed a second time after the removal of outlier pixel values. 
The approximate wavelength solution is finally obtained interpolating the
wavelength values versus the corresponding pixel values  with a low-order 
polynomial (degree=3)

\subsection{Wavelength dispersion solution for a single frame}\label{sec:wavel-disp-solut}

Now that an approximate solution is available, the determination of
the final dispersion solution is an iterative process where the
resulting precision is increased with each iteration. 
These are the steps that we followed:
\begin{itemize}
 \item at each iteration more faint lines are measured in the
   calibration frame and associated with lines in the reference
   spectrum.
\item Lines with \fwhm\ higher then a given threshold, nominally twice
  the instrumental \fwhm\ at that wavelength, are excluded from the list;
\item The dispersion solution is computed again, and the order of the
  polynomial can be increased according to the number of added lines.
\item Lines that have an interpolated wavelength that differs from the
  reference value over a given threshold are rejected and not included
  in next iterations; the threshold is initially very large and it
  is rescaled at each iteration.
\end{itemize}

\begin{figure}
\includegraphics[width=\columnwidth]{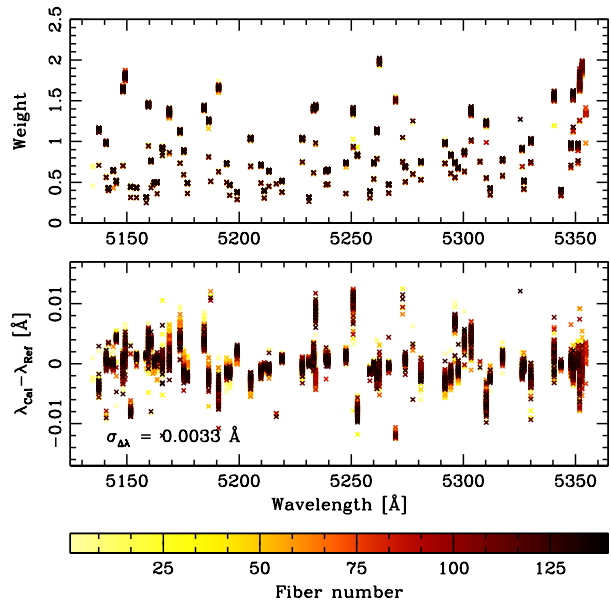}
\caption{Two quantities associated with pixel-wavelength calibrations
  are plotted as functions of wavelength.
  In the upper panel, the weight associated to each line is shown.
  In the lower panel, the difference between the wavelength determined
  using the dispersion solution for the measured center of the line,
  and the reference wavelength from CES spectra is
  plotted. The standard deviation of this difference
    is reported in the figure. Points from different fibres are
    color-coded accordingly.}
\label{fig:wflat_calib}
\end{figure}

In the last two iterations, the lines are weighted according to their
brightness and their width in both the calibration spectra and the
reference one.  
After several tests Equation~\ref{eq:wflat_calib_weights}
was found to provide the best combination of these parameters,
\begin{equation}\label{eq:wflat_calib_weights}
w_i=\left ( \sqrt{\sigma_{cal,i}^2 + \sigma_{ref,i}^2 } +
500/\sqrt{A_i} \right )^{-1}
\end{equation}
where $w_i$ is the weight associated to the line $i$,
$\sigma_{cal,i}$ and $\sigma_{ref,i}$ are the widths of the line 
in the calibration and reference Th-Ar spectrum respectively,
and $A_i$ is the measured amplitude of the line. An example of the
weight association as a function of wavelength is given in the upper
panel of 
Fig.~\ref{fig:wflat_calib} for a randomly selected calibration
frame. 

In the last two iterations, the polynomial function is substituted with a
piecewise polynomial (in the form of Basis splines or B-splines). To
fit the curve to our data we made use of the EFC routine included in
the SLATEC Common Mathematical Library\footnote{
The official repository of
  Version 4.1 ( July 1993) is \url{http://www.netlib.org/slatec/}}, a
collection of mathematical and statistical routines written in Fortran 77.
This procedure is repeated individually for each fibre. 

In the lower panel of Fig.~\ref{fig:wflat_calib} the
difference between the wavelength obtained with the dispersion
solution (from the position in the pixel space) and the reference
wavelength from the CES spectrum is shown as a function of
wavelength. The standard deviation of this difference, determined using all the
lines measured in a calibration frame,  is on average $\sigma=0.0033$
\AA, which corresponds to $\simeq 1/15^{\textrm{th}} $ of a pixel. 

\subsection{Overall calibration}\label{sec:overall-calibration}

After the wavelength calibration of a frame, the lines used for the
dispersion solution are stored. 
Our dataset span a large temporal range, from almost the beginning of
the scientific observations of GIRAFFE in 2003 until 2009. The
information on emission Th-Ar lines from all these frames is
collected in order to identify a stable set of lines and improve the
dispersion solution. Only the lines that
have been successfully measured on $90\%$ of the fibres are retained.
The only exception is for those lines that are closer than $10$ \AA\ to
the edge of the spectrum, for the reasons explained in
\S\ref{sec:first-appr-wavel}.
The weight associated to each line is the mean of the weights calculated
in each calibration fibre 
using Equation \ref{eq:wflat_calib_weights}.
A total of 114 lines has been selected (Table~\ref{tab:wl_wg_lines}).
The dispersion solution is then computed again, using the same
function of the last iteration in \S\ref{sec:wavel-disp-solut} but with
the improved line list. 

\begin{table}
\center
\caption{Lines and Weights used for the Dispersion Solutions}
\label{tab:wl_wg_lines} 
{\scriptsize
\begin{tabular}{cccccc}
\hline
n &  Wavelength [\AA] & Weight & n &  Wavelength [\AA] & Weight\\
\hline\hline
   1  & 5134.7457 & 8.776  &    58  & 5239.5502 & 8.574 \\ 
   2  & 5136.1202 &26.557 &     59  & 5241.8563 &30.608 \\ 
   3  & 5137.4775 &15.334 &     60  & 5243.7542 &24.384 \\ 
   4  & 5140.7642 &13.070 &     61  & 5247.6542 & 9.685 \\ 
   5  & 5141.7838 & 5.382 &     62  & 5250.8742 &18.822 \\ 
   6  & 5143.9137 & 8.399 &     63  & 5252.8043 &11.150 \\ 
   7  & 5145.3047 & 6.584 &     64  & 5254.2542 &12.568 \\ 
   8  & 5148.2122 &19.641 &     65  & 5258.3603 & 4.827 \\ 
   9  & 5149.2102 &21.811 &     66  & 5260.1046 & 9.866 \\ 
  10  & 5151.6104 & 5.564 &     67  & 5261.4763 &15.428 \\ 
  11  & 5154.2434 & 5.420 &     68  & 5262.6055 &23.083 \\ 
  12  & 5157.4512 &26.060 &     69  & 5264.8003 &19.409 \\ 
  13  & 5158.6042 & 3.818 &     70  & 5266.7102 & 6.025 \\ 
  14  & 5159.5911 &19.307 &     71  & 5269.7842 &19.385 \\ 
  15  & 5160.7159  &9.8942 &     72  & 5272.9343 &14.107 \\ 
  16  & 5161.5402 & 4.598 &     73  & 5274.1183 & 9.093 \\ 
  17  & 5162.2882 & 6.315 &     74  & 5277.4967 &13.711 \\ 
  18  & 5163.4582 & 6.360 &     75  & 5281.0698 & 9.834 \\ 
  19  & 5164.4022 &23.736 &     76  & 5282.4010 &23.019 \\ 
  20  & 5165.7682 &12.135 &     77  & 5283.6902 &24.195 \\ 
  21  & 5166.6522 &23.825 &     78  & 5286.8895 &15.986 \\ 
  22  & 5168.9242 &18.375 &     79  & 5289.9048 &21.810 \\ 
  23  & 5170.2482 &23.561 &     80  & 5291.8202 &13.201 \\ 
  24  & 5170.2482 &25.057 &     81  & 5294.3981 &11.068 \\ 
  25  & 5173.6782 &14.890 &     82  & 5295.0866 &17.948 \\ 
  26  & 5175.3262 &11.583 &     83  & 5296.2823 & 9.744 \\ 
  27  & 5176.9642 & 6.212 &     84  & 5297.7575 & 8.974  \\
  28  & 5178.4602 &21.545 &     85  & 5300.5243 &11.599 \\ 
  29  & 5180.7198 &22.175 &     86  & 5301.4042 &13.648 \\ 
  30  & 5182.5242 &15.943 &     87  & 5303.4689 &18.513 \\ 
  31  & 5184.4562 &19.374 &     88  & 5305.6923 &16.192 \\ 
  32  & 5186.4122 &16.782 &     89  & 5307.4660 &14.933 \\ 
  33  & 5187.3402 & 5.066 &     90  & 5309.6143 &20.676 \\ 
  34  & 5190.8742 &19.680 &     91  & 5310.2690 &16.136  \\
  35  & 5194.4582 & 9.513 &     92  & 5312.0023 & 5.200 \\ 
  36  & 5195.8142 & 5.877 &     93  & 5315.2242 &19.243 \\ 
  37  & 5197.2362 &21.669 &     94  & 5317.4943 &10.140 \\ 
  38  & 5199.1702 & 4.668 &     95  & 5320.7727 &21.339 \\ 
  39  & 5202.0231 &21.904 &     96  & 5322.9002 &20.316 \\ 
  40  & 5203.8475 &19.515 &     97  & 5325.4079 &17.428 \\ 
  41  & 5205.1522 &13.871 &     98  & 5326.2724 &11.946 \\ 
  42  & 5206.5122 &19.620 &     99  & 5326.9763 & 6.434 \\ 
  43  & 5207.7907 &24.472 &    100  & 5329.3943 &10.258 \\ 
  44  & 5209.7252 & 9.314 &    101  & 5330.0782 &13.466 \\ 
  45  & 5211.2309 & 4.933 &    102  & 5333.3903 &24.130 \\ 
  46  & 5213.3589 & 8.251 &    103  & 5337.0125 &24.140 \\ 
  47  & 5216.6123 & 6.105 &    104  & 5338.2680 &23.143 \\ 
  48  & 5219.1121 & 6.626 &    105  & 5340.5048 &19.221 \\ 
  49  & 5220.9031 & 8.875 &    106  & 5343.5817 & 4.772 \\ 
  50  & 5226.5323 &24.091 &    107  & 5345.3154 &22.528 \\ 
  51  & 5228.2291 &14.010 &    108  & 5346.3805 &20.157 \\ 
  52  & 5228.9985 &22.407 &    109  & 5347.9737 &12.361 \\ 
  53  & 5231.1600 & 3.836 &    110  & 5349.0053 &19.482 \\ 
  54  & 5233.2263 &18.568 &    111  & 5351.1289 &12.436 \\ 
  55  & 5234.1222 &19.372 &    112  & 5351.8442 &21.831 \\ 
  56  & 5237.9123 &28.873 &    113  & 5353.0208 &22.987 \\ 
  57  & 5238.8142 & 8.417 &    114  & 5354.6055 &17.678 \\ 
\hline 
\end{tabular}
}
\end{table} 

\begin{figure}
\includegraphics[width=\columnwidth]{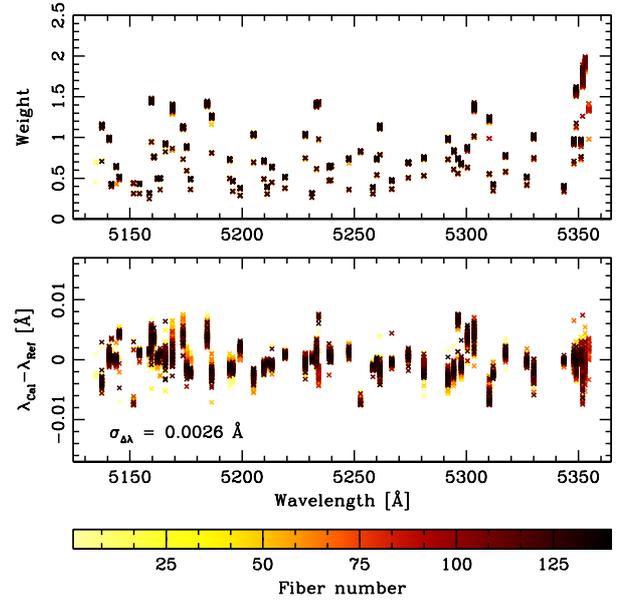}
\caption{As in Fig.~\ref{fig:wflat_calib}, on the same
  calibration frame but using the refined line list.
  Note the reduction in the standard deviation compared to the previous
  figure.}
\label{fig:wflat_calib_all}
\end{figure}

Using the same set of lines and weights to
compute the dispersion solution results in a more homogeneous
calibration of the entire dataset, while using the more stable lines
reduce the scatter around the computed dispersion.  
In Fig.~\ref{fig:wflat_calib_all}, the resulting overall calibration
for the same frame in Fig.~\ref{fig:wflat_calib} is shown.  
A further reduction of the dispersion from $\sigma=0.0033$ \AA\ to
$\sigma=0.0026$ \AA\ 
of the difference between measured and reference wavelength of
individual lines is obtained. 
To have an idea of the goodness of the dispersion 
solution, this result can be compared with the average pixel size of 
$\simeq$0.05~\AA,  or typically $\lesssim 1/19^{th}$ of the size of a pixel.

 \section{SimCal calibration}\label{sec:simcal-calibration}

A simultaneous calibration lamp provides a reference system obtained
simultaneously with the observations, to take into account drifts of the
spectrum in the wavelength space that may occur during the time between the
Th-Ar lamp calibration frame and observations. 

GIRAFFE has five fibres available for Th-Ar SimCal, homogeneously
distributed between the science fibres. 
In the ESO pipeline, drifts are corrected by cross-correlating the SimCals
with the provided numerical mask; the resulting offsets are linearly
interpolated across the CCD in the cross-dispersion direction to
determine the wavelength offset to be applied for each fibre to the dispersion
solution obtained using the day-time calibration frame.
Two apparently reasonable assumptions are made here: the drift is constant 
in the velocity
space across the wavelength range of the considered order, and that
the overall shift is linear in the spatial direction of the CCD. 

We first check if these two assumptions are correct. 
In order to do so, we need to compare two Th-Ar
  calibration frames with a measurable drift. 
We then opted to compare calibration frames from two different nights.
Emission line positions for each fibre are measured in 
both frames and the difference in pixel space is taken. 
Conversion of differences from pixel space $\Delta x_i^{pixel}$  into
radial velocity space $\Delta RV_i$ for the $i_{th}$ line
is given by:
\begin{equation}\label{eq:pixel_2_rv}
\Delta RV_i =  \Delta x_i^{pixel} \delta \lambda(i) \ c / \lambda_i
\end{equation}
with $\lambda_i$ given by the day-time wavelength dispersion solution,  
$\delta \lambda_i$ is the size of the pixel in the wavelength space at the
position of the $i_{th}$ line.

\begin{figure}
\includegraphics[width=\columnwidth]{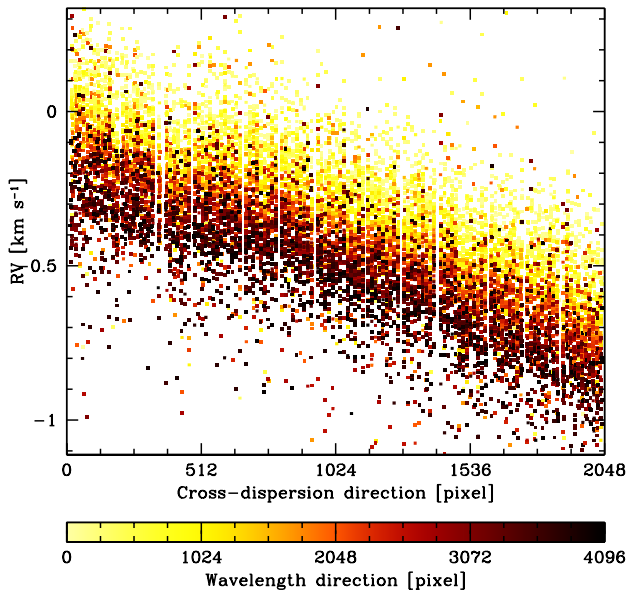}
\caption{Radial velocity shifts between the
    calibration frames of two typical nights, along the cross-dispersion direction. 
  The position along the wavelength dispersion direction is
  color-coded. This plot shows that the common practice of assuming a
  constant value along one axis is not justified.
\label{fig:simcal_crosslin_x}}
\end{figure}

\begin{figure}
\includegraphics[width=\columnwidth]{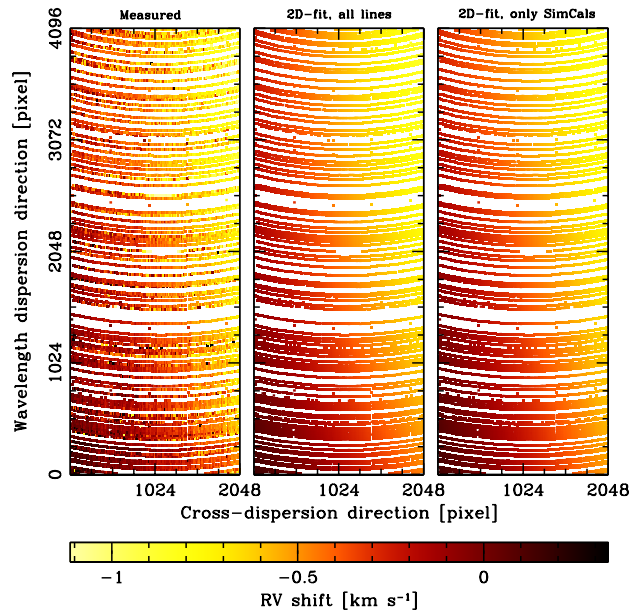}
\caption{First panel: Radial velocity shift of Th-Ar emission lines  between the calibration
  frames of two different nights. 
  Second panel: 2-D polynomial fit to the data of the 
  first panel, using all the available fibres in the frame.
  Third panel: the same
  fit is performed using only the Simultaneous Calibration
  fibres. Radial velocity shifts are color-coded. The
  chosen model is able to reproduce the global shape of the RV shifts
  while using the SimCals alone, with little influence from outliers.}
\label{fig:simcal_cheby}
\end{figure}

Fig.~\ref{fig:simcal_crosslin_x} and the first panel of Fig.~\ref{fig:simcal_cheby} show the difference between 
the calibration frames of two typical nights.
Different night combinations show similar results, with the only significant 
difference being the amount of overall radial velocity shift. 
It appears clear that both ESO pipeline assumptions here do not hold: 
the shift is not constant for a single fibre, and it is not a linear 
function of the cross-dispersion direction. 
Probably these differential shifts also happen during individual 
nights, even if in a less pronounced way. 
We conclude that correcting for a single value of the radial velocity 
will not provide a good correction for drifts, specially when temperature and air pressure inside the spectrograph are not actively  
controlled, as in our case. 

In our new analysis, a 2D polynomial fit (Chebyshev polynomials of the first 
kind) of degree $(2,2)$ has been chosen to model the shift in radial velocity
as a function of pixel coordinates.
However, during science observations only five fibres out of 132 are available
for simultaneous calibration (fibres 1, 32, 63, 94 and 125). 
To check if the polynomial fit works as well with a reduced number of 
fibres, we performed two fits of the measured RV differences for two given 
nights (first panel of Fig.~\ref{fig:simcal_cheby}): the first 
one using all the available fibres (second panel of
Fig.~\ref{fig:simcal_cheby}), and the second one using only the
SimCals fibres (third panel of Fig.~\ref{fig:simcal_cheby}), in both
cases using the same function to fit the measured differences.  

\begin{figure}
\includegraphics[width=\columnwidth]{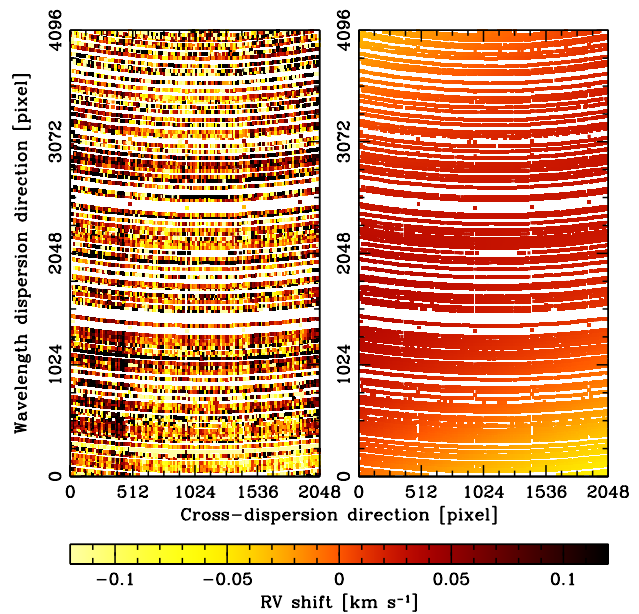}
\caption{First panel: differences between the SimCals drift model and the
  observed radial velocity shift. Second panel: differences between the
  SimCals drift model and the same model using all the calibration
  fibres. The same calibration frames of
  Fig.~\ref{fig:simcal_crosslin_x} and \ref{fig:simcal_cheby} have
  been used. The scatter in RV of individual lines is clearly larger
  than the error we introduce in the global determination of the drift
by using only the SimCals fibres.}
\label{fig:simcal_diff}
\end{figure}

In the example shown in Fig.~\ref{fig:simcal_diff}, the difference
between the drift model using only SimCal fibres and measured line
positions (left panel) has a mean of 
$\langle {\Delta _{RV}} \rangle$~$<$~10~\ms\ and standard deviation 
$\sigma \simeq 125$~\ms\ after $5-\sigma$ clipping. 
The difference with the model obtained using all the fibres
(right panel of Fig.~\ref{fig:simcal_diff}) shows a systematic of
$\langle {\Delta _{RV}} \rangle$~$<$~10~\ms\ and a 
$\sigma \simeq 20$~\ms\ (after $5-\sigma$ clipping). 
Considering that we are using only 5 fibres out of 132 to redetermine 
the drift across the whole detector, this is a reasonable result.  
Finally, we note that polynomials with higher degrees 
resulted in fit too sensitive to outliers when using only the five 
SimCal fibres.

 \section{Radial velocity determination}\label{sec:radi-veloc-determ}

In next section the three most popular techniques in radial
velocity measurement are briefly described:
 \textit{Classical} Cross-Correlation Function (CCF), Numerical CCF (nCCF), and Synthetic Template
Matching. 
The steps for determination of radial velocities 
for each fibre within each exposure are then described in detail.

 \textbf{Classical Cross Correlation function:}
This is the technique developed by \cite{Tonry:1979aa} and implemented
in the RVSAO package \citep{Kurtz:1998aa}\footnote{A detailed
  description of the package is available at \url{http://tdc-www.harvard.edu/iraf/rvsao/xcsao/xcsao.proc.html}}.
The observed spectrum is rebinned into a linear scale
in logarithmic wavelength, continuum
normalized and then the Fourier power spectrum (FPS) is computed. 
The Fourier power spectrum of a pre-selected
template spectrum is computed too, and the two FPS are
cross-correlated to determine the radial velocity shift. In some
modifications of the algorithm, the CCF is computed using the spectra
directly, in the log-wavelength space.

\textbf{Numerical CCF:}
This is the technique introduced by \cite{Baranne:1996tb} and
improved by \cite{Pepe:2002ab}, also known as \textit{CORAVEL-type} CCF. 
The observed spectrum is cross-correlated in wavelength space with a
numerical mask with non-negative values at the central position of the
line, and zero otherwise. 
Each value of the mask, or \textit{weight}, is determined using a synthetic
spectrum, taking into account the depth of the line and its width, in
order to optimize the extraction of the doppler shift information. 
The cross correlation function is constructed by shifting the mask as a
function of the Doppler velocity, and integrating the product of the
mask with the observed spectrum.  
As a result of this definition, the value of the RV is given by the minimum in
the cross-correlation function. To avoid confusion we will refer to this definition as \textit{nCCF}
This technique has proven its reliability in the discovery of several 
exoplanets, and achieves best results when a large number of lines is 
available (\eg, \citealt{Pepe:2004hi}).

 \textbf{Synthetic Template matching:}
A synthetic spectrum to be used as rest-frame reference is generated using 
the atmospheric stellar parameters from photometry. 
The observed spectrum is shifted in the RV space and for every shift
$v_r$ the $\chi^2$ of the difference between the observed and the
synthetic spectrum is computed. 
The value of $v_r$ 
that gives the lowest $\chi^2$ is taken as the radial velocity of the star.

The main disadvantage of Classical CCF technique is
  that the rebinning process (required for the  FPS computation) can introduce correlated
noise in cases of very low \snr\ spectra. Since we want to avoid
spectra rebinning in the first place this
technique and further derivation have not been considered in the
following analysis.

\subsection{Application of the numerical techniques}\label{sec:math-applications}
We describe now our implementation of the Synthetic Template matching technique.
To begin, the sky spectrum is removed from the observed one and the 
result is normalized to the continuum level of the stellar flux as later
described in \S\ref{subsec:cont_sky_norm}.  
Instead of a  $\chi^2$ function, we minimize the 
least-square function \textit{LSQ}
defined in Equation~\ref{eq:fitness_function}, where $f_{\star}$ 
is the observed stellar spectrum (after the steps described above),
$f_{\textrm{syn}}$ is the synthetic one, and the sum is over the
$n_{pixel}$ data points of the observed spectrum.
Pixels affected by cosmic rays or CCD defects and spectral ranges
contaminated by Th-Ar emission from close SimCals fibres are flagged
with a zero value in $mask$, and unitary value otherwise. 

\begin{equation}
LSQ(v_r) = \sum_i ^{n_{pixel}} \left ( \frac{ f_{\star}(i) - f_{\textrm{syn}}(i,-v_r)
}{1.2 +f_{\textrm{syn}}(i,-v_r) } \right ) ^2 mask(i).
\label{eq:fitness_function}
\end{equation}

For each determination of the $LSQ$ function the synthesis is
shifted by $-v_r$ in the radial velocity space (in the opposite
direction since $v_r$ is the velocity associated to the star) and then 
rebinned into the wavelength scale of  $f_{\star}$. 
This approach avoids the more problematic rebinning of the noisier 
observed spectrum, and is the reason why the dependence on $v_r$ has been 
included in the synthetic spectrum $f_{\textrm{syn}}(i,-v_r)$.

The denominator of Equation~\ref{eq:fitness_function} differs
significantly from the one of a standard $\chi^2$ function. 
We did not include the dependence on the continuum in $LSQ$
determination because this shape is changing with time and
changes in instrumentation (\eg\ the change of CCD in May
2008\footnote{See the FLAMES Giraffe
Data Processing and Quality Control website \url{http://www.eso.org/observing/dfo/quality/index\_giraffe.html}}); thus
it could have introduced a bias depending on the epoch of observations. 
We instead decided to divide by the synthetic spectrum
to give more weight to spectral lines, while a constant has been
added to avoid an excessive weight for the deepest lines.
The value of the additive constant has been found
  empirically by performing several tests on a subsample of stars of
  different spectral kind and with
  multiple observations.

\begin{figure}
  \includegraphics[width=\columnwidth]{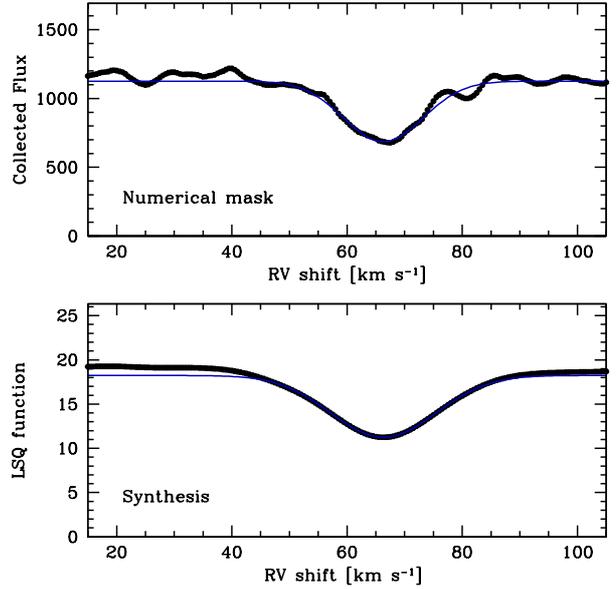}
\caption{Comparison between the ``CORAVEL-type'' Numerical CCF (upper panel) 
   and Synthetic Template Matching (lower panel) 
   on the spectrum of a SGB star with $V=16.4$.
  The blue line represents Gaussian fits to these functions.}
\label{fig:CCF_compare}
\end{figure}

We have tested the Numerical CCF and the Synthetic Template
Matching on several stars with spectra at various \snr\ to choose
the technique that better suits our dataset. The
resulting functions for a SGB star with $V=16.4$ are shown in
Fig.~\ref{fig:CCF_compare} in order to show the differences
between the two techniques. For a given value of the radial velocity
the Numerical CCF technique is using only the portions of the spectrum
that are falling inside the positive values of the mask (namely
$\delta-$functions), while with the Synthetic Template Matching 
the $LSQ$ function is determined over the entire spectrum. In both
cases the width of the resulting functions will be given by the sum in
quadrature of the average width of the spectral lines in the observed
spectrum and the width of the lines in the mask/synthesis, hence with the
Synthetic Match technique the final functions will be broader with respect to the
Numerical derived one. A broader function will result in a greater
uncertainty in the determination of its center, \ie\ the Synthetic
Match delivers a less precise radial velocity than the CORAVEL
technique (if the same linelist is used in both cases). On the other
hand, while this may be
true in an ideal situation, when the limited wavelength range of the
spectra allows to use only a few hundred  of lines (with respect to the thousands of lines as in
\citealt{Pepe:2004hi}) as in our case, the CORAVEL
technique is too sensitive to noise, as exemplified in
Fig.~\ref{fig:CCF_compare}. We therefore decided to use the
Synthetic Match to determine the RV of the whole set of stars.

\subsection{Continuum normalization and Sky flux determination}\label{subsec:cont_sky_norm}

Continuum normalization is one of the most daunting tasks in stellar
spectroscopy. 
Generally speaking, there are several ways to deal with the continuum. 
When a spectrum has extended wavelength coverage as in
the case of Echelle spectra, and the aim is to fit synthetic spectrum
to the data, the continuum is obtained by fitting pre-selected feature-less
spectral regions around the one to be analysed (as done by 
\citealt{Kirby:2008cr}, but many other examples are available in literature). 
In other cases as in the case of Equivalent Width measurements the 
interest is focused on individual lines and it is possible to use those 
surrounded by at least a few \AA ngstroms of local
continuum (see for example the criteria used by 
\citealt{Sousa:2007gn} to create their linelist).

Our dataset however presents several major complications. 
The spectral range is very short, $\Delta \lambda = 214$~\AA, and 
very crowded, so there are no line-free regions
available to set a continuum level reliably.
In fact, the HR9 setting was chosen for the richness in this spectral window.  
The wings of the Magnesium-I triplet  ($5167.33$ \AA, $5172.70$ \AA,
$5183.62$ \AA) affect nearly one quarter of the spectrum beginning
at the low-wavelength edge of our spectra, so that the continua must be 
extrapolated from the last three quarters of the spectra. 
Moreover, many lines are blended or very close, making the determination 
of local continua very difficult.

To determine the continuum for each spectrum, we used a 
technique similar to the one introduced in LM14.
Briefly, the continuum level and sky contribution are determined 
by using a spectral synthesis, obtained using the photometric stellar
parameters and average metallicity of the cluster, and the Solar Flux
Atlas \citep{Kurucz:1984ab} as a template sky spectrum, both normalized
to unity. 
Photometric atmospheric parameters for M\,4 have been derived using the 
photometric catalog kindly provided by Peter B.~Stetson \citep{DAntona:2009kg}
and the calibrations from \cite{Ramirez:2005bz} and \cite{Casagrande:2010a}. 
The latest determinations for reddening and distance from 
\cite{Hendrick:2012cw} have been adopted:
\ $E(B-V)= 0.37 \pm 0.01$, $R_V=3.62$ and
$D_{\rm cluster}=1.80 \,{\rm kpc}$.  
Syntheses are generated using the current version of the Local 
Thermodynamic Equilibrium (LTE) code {\sc MOOG} \citep{Sneden:1973el}, 
and \cite{Kurucz:1992ab} atmosphere models calculated with the $\alpha
-$enhanced New 
Opacity Function Distribution (\textrm{AODFNEW}, \citealt{Castelli:2004ti}). 
More details on the photometric parameter determination, atomic line 
parameters and synthesis calculation can be found in LM14.

To apply this technique, we need to know the radial velocity of the star. 
At this stage high precision in RV is not required. 
A single pixel has an average size of $2.5 \,\kms$, so an 
error of several hundreds \ms\ still yields a good fit of 
the synthesis to the observed spectra, given the fact that our goal at 
this stage is to determine the continuum level and not to derive accurate
radial velocities or stellar parameters.
The radial velocity is derived using a numerical CCF \citep{Baranne:1996tb} 
with line positions and weight derived from the same synthetic spectrum 
that will be used to determine the continuum, following the
prescriptions in \cite{Pepe:2002ab}.  

\begin{equation}\label{eq:ccf_definition}
nCCF(v_r) = \sum^i nCCF^i(v_r)=\sum^i \int_{\lambda^i_{v_r} - \delta
  \lambda/2}^{\lambda^i_{v_r} +
\delta \lambda/2} f(\lambda) c^i d\lambda .
\end{equation}

As can be seen from Equation~\ref{eq:ccf_definition}, the value of
the nCCF at a given radial velocity point $v_r$ is proportional to
the sum of the stellar flux collected by each $\delta$
  function
 $i$ in the
numerical mask, rescaled for the weight $c_i$. 
In this equation, $\lambda^i_{v_r}$ is the wavelength of the line $i$ 
shifted for the radial velocity of the nCCF and
$\delta \lambda$ is the size of the hole in the numerical mask. 

\begin{figure}
\centering
\includegraphics[width=\columnwidth]{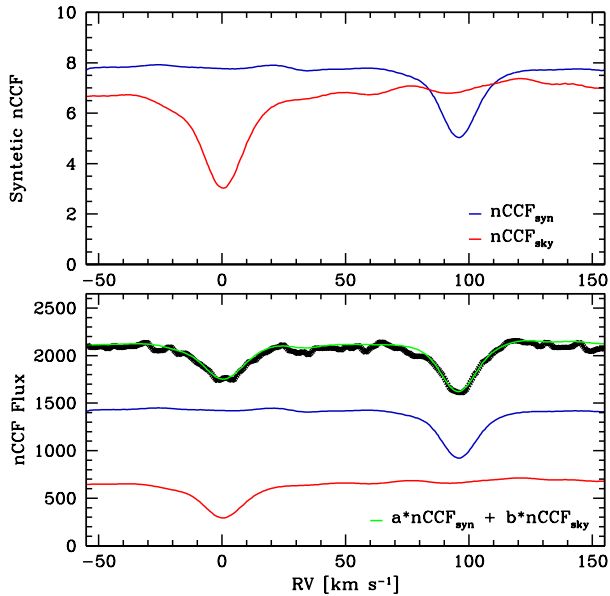}
\caption{In the top panel, the nCCF from the normalized spectrum of the
  synthesis (blue line) and the sky (red) are shown. 
  In the bottom panel, $nCCF_{syn}$ and $nCCF_{sky}$ are rescaled according 
  to the measured flux to show the relative influence, and their sum 
  (green) is compared with the observed nCCF (black). 
A star with $V_{mag}=16.4$ is considered here as example.
} 
\label{fig:ccf_starsky_flux}
\end{figure}

The result of the numerical CCF allows us to determine the star/sky flux ratio. 
The nCCF is computed for the smoothed synthetic spectrum ($nCCF_{synth}$)
and the sky spectrum ($nCCF_{sky}$) without any further correction and using
the same numerical mask. 
The cross-correlation function of the observed spectrum $nCCF_{star}$ is then 
fitted with a linear combination of $nCCF_{synth}$ and $nCCF_{sky}$, 
being the linear coefficients of the fractions of stellar and sky
fluxes.

The upper panel of Fig.~\ref{fig:ccf_starsky_flux} shows
the $\rm{nCCF}_{synth}$ (red line) and the  $\rm{nCCF}_{sky}$ (blue line) derived 
from normalized spectrum. 
The sky spectrum has a lower continuum and a deeper nCCF because of the 
higher metallicity and lower temperature respect to the observed star. 
In the lower panel, the two nCCFs are rescaled according
to their estimated fluxes in order to show their relative influence, and 
their sum (green line) is compared with the observed nCCF (black points).  
Note here that this technique works best when the
observed RV of the star (including Earth motion contribute) differs
significantly from the RV of the sky, \ie\ $|v_r^{star}| > 20$~\kms.

Both solar and synthetic spectra then are rebinned into the wavelength 
scale of the science spectrum (keeping constant the flux for wavelength 
unit) and degraded to match the spectral resolution of the instrument.
Finally the instrumental effects (\eg flat-fielding) are applied to make the comparison
with the observed spectrum possible. 
A normalized linear combination of the two modified spectra, with the 
sky/flux ratio as coefficients, is multiplied by a polynomial function: 
its coefficients are determined by fitting the observed spectrum. 
The partial continuum levels of star and sky are obtained by substituting 
the initial synthesis and the sky spectra respectively with unitary spectra 
and applying all these same steps, with the exception of the fit of the 
polynomial function which coefficients are now known. 
The overall continuum level is the linear combination of the two partial 
continua. 

When sky subtraction and continuum normalization play a decisive role, 
\eg\ in the determination of the stellar atmosphere parameters, more 
refined values for the star/sky ratio and continuum coefficients 
should be obtained as described in LM14,
where a much larger emphasis was given to the quality of the
sky subtraction and continuum normalization. In that case, the
star/sky ratio was fitted simultaneously with the stellar parameters of the
synthetic spectra and the continuum placement, while here it is
determined prior to the RV measurement.

Atmospheric parameters and abundance measurements are particularly
sensitive functions of the depth of the spectral lines, \eg\
underestimating the continuum level would result in lower abundances,
so refined values for the star/sky ratio and continuum coefficients
are required. Compared to this, measuring the shift in wavelength of a
spectrum
is a simpler process. Thus while removing the sky contribution and
removing instrument signatures can help,
ultimately the achievable precision will depend on other factors such
as the quality of the wavelength calibration and the \snr\ of the
spectra. For this reason we have decided to stick with the simpler
approach presented here. 

\begin{figure}
  \includegraphics[width=\columnwidth]{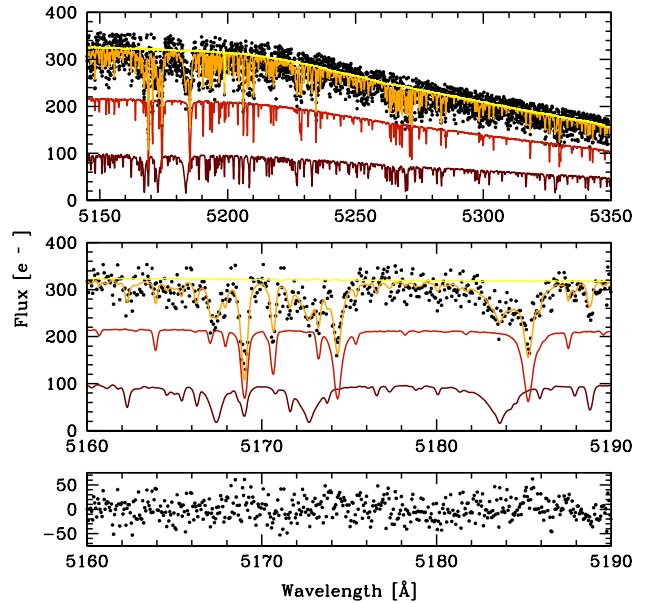}
\caption{
  Upper panel: the contribution of sky (brown line) and star
  (red line), their combination (orange line) and the derived normalization 
  level (yellow line) compared to the observed spectrum (black points) 
  for a program star with $V_{mag}=16.4$. 
  Middle panel: the same spectra zoomed in on the Magnesium triplet
  region. Lower Panel: the residuals of the fit (observed spectrum
  minus green line) have a standard deviation of 21 $e^{-}$,
  consistent with the \snr\ of the spectrum.}
\label{fig:cont_example}
\end{figure}

Fig.~\ref{fig:cont_example} shows the continuum determination 
for a $V_{mag}=16.4$ star, with photometrically derived 
\teff~=~6000~{\rm K}, \logg~=~4.0, and \gfeh~=~$-$1.15.
The spectrum was taken on 2006 September 04, about five days before full Moon,
at which time 
the angular separation between M4 and the moon was $\simeq 40^o$.
Extraction of a pure stellar spectrum here is very  
difficult:
the \snr\ is low, the sky contamination is large, the stellar lines are 
strong, the Mg I triplet in star and sky spectra severely depress the 
fluxes at the blue end of the spectral range, and as always the total 
spectral range is short.
All these effects combine together to make it almost impossible to find 
suitable line-free continuum zones to normalize the spectra. 
However, as we show in the lower panel of Fig.~\ref{fig:cont_example}, 
our technique works well even in these difficult cases. 
The observed spectrum is closely matched by the combination (green
line) of the contribute of the sky (red line) and the stellar spectrum (blue
line), with a sky/star ratio of 0.46.
The knowledge of the individual contributes allows a reliable
continuum determination even in problematic region of the spectra, 
such as around the Magnesium triplet.
The residuals of the fit in this region (in the lower panel) have a standard deviation
  of 21 $e^{-}$, which is very close to the value of the photon noise,
  $\simeq 17 e^{-}$.

For faint stars, even if observed only a few days after new moon, the
sky can contribute up to half of the observed spectra of faint M\,4 targets.
Otherwise the additional lines would lower the continuum
level determined using only the stellar synthesis. 
Fortunately, our spectral range ($5140$\AA\ $< \lambda < 5360$\AA) 
is free from atmospheric absorption lines,
so no correction is required for this potential issue.

\begin{figure}
\centering
\includegraphics[width=\columnwidth]{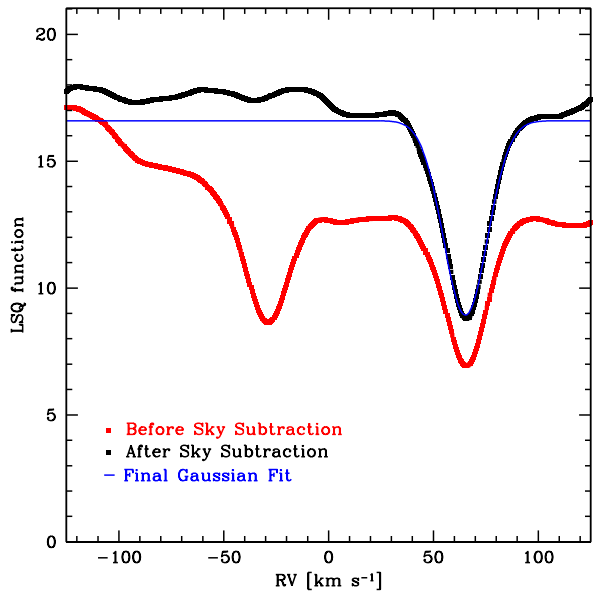}
\caption{The $LSQ$ function for the same observations shown previously, before 
   (red line) and after (black) sky subtraction. 
   The blue line represents
   the gaussian fit to determine the RV of the star.} 
\label{fig:sky_subtraction}
\end{figure}

Fig.~\ref{fig:sky_subtraction} shows the $LSQ$ function for this observation
before and after the removal of the sky spectrum. 
The spectrum has been normalized using the continuum associated to 
the combined sky$+$synthesis in the first case, and the continuum associated 
to the synthesis alone in the second case after sky removal.
The slope introduced by the contribution of the sky, which
inevitably alters the shape of the curve associated to the star, is
essentially gone after sky-subtraction from the observed spectrum. 
Finally, confidence in this approach for removing sky
contamination and continuum normalization is increased
by noting that stars with multi-epoch observations show
lower internal dispersion in their RVs than can be obtained with other
methods.

 \section{Results}\label{sec:results-discussion}

After sky subtraction and normalization, the radial velocity is
finally determined with a gaussian fit of the resulting $LSQ$ function,
determined as described in \ref{sec:radi-veloc-determ}.

Radial velocities shifts due to the intrinsic velocity of
the star and the Earth motion are then applied to the stellar synthetic
spectrum. 
Heliocentric and Barycentric corrections are determined
using the \textrm{FORTRAN} subroutine by \cite{Stumpff:1980ab}.

The error in the radial velocity associated with a single exposure is a
function of the spectral type of the star, the characteristics of the
instrument, the radial velocity determination technique and the \snr\
of the spectrum.  
Instead of trying to determine a formal velocity error of a single exposure, 
we determine an empirical error using the several radial velocity
measurements available for each star.

The weighted mean velocity of a star $v_r$ is determined using 
Equation~\ref{eq:rv_wmean}, where $r$ is the index of the star, $j$ is the
index referring to individual exposures and $n$ is the
total number of observations available for the given star. 
\begin{equation}\label{eq:rv_wmean}
\overline{v_r} = 
\frac{ 
   \Sigma_{j=1}^{n} w_{r,j} v_{r,j}
}{
   \Sigma_{j=1}^{n} w_{r,j}
}.
\end{equation}
We use as weight $w_{r,j}$ the inverse of the minimum of the $LSQ$
  function (Fig.~\ref{fig:sky_subtraction}) after normalizing it for
the number of pixels used. 

\begin{figure}
\includegraphics[width=\columnwidth]{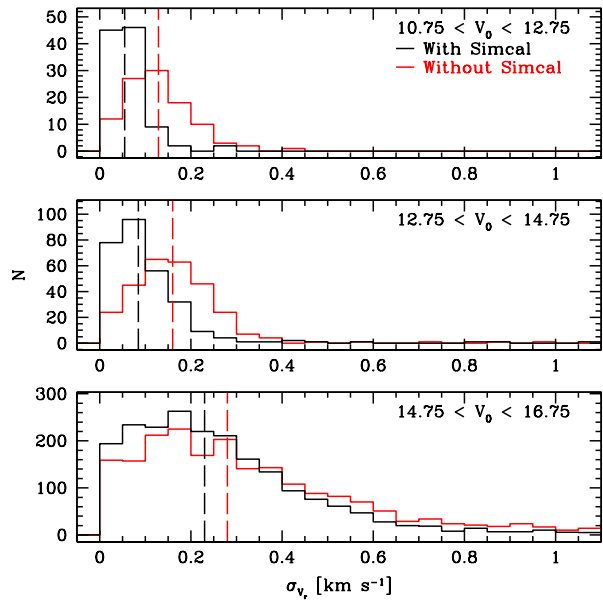}
\caption{Distribution of the radial velocity dispersion $\sigma_{v}$
  associated to each star before (red lines) and after (black lines)
  the SimCals correction. Vertical lines mark the
    median values of the two distributions.
  Three ranges in de-reddened V magnitude are considered.}
\label{fig:RV_rms_compare}
\end{figure}

Following 
S09, the unbiased estimator for the
population variance is determined with  
\begin{equation}\label{eq:rv_wsigma}
(\sigma_{v_r})^{2} =
\frac{ \Sigma_{j=1}^{n} w_{r,j} }{(\Sigma_{j=1}^{n} w_{r,j})^{2} -
  \Sigma_{j=1}^{n} w_{r,j}^{2} }
~ \Sigma_{j=1}^{n} w_{r,j} (v_{r,j} -  v_r)^{2}.
\end{equation}
It is interesting to compare the distribution of the radial velocity
dispersion associated to each star before and after the SimCals correction 
for three different intervals in de-reddened \textrm{V} magnitude
(Fig.~\ref{fig:RV_rms_compare}). 
Going from the lower panel to the upper one, we can see how the correction 
for instrumental drifts in the radial velocity becomes increasingly 
more important for brighter stars, \ie\ with higher \snr. 
This improvement is highlighted by the difference in the median  values of the two distributions (vertical lines in the Figure).
Although radial velocity measurements for faint stars are mostly
dominated by photon noise, a small improvement in
 the $\sigma_{v_r}$ is still present.

\begin{figure}
\includegraphics[width=\columnwidth]{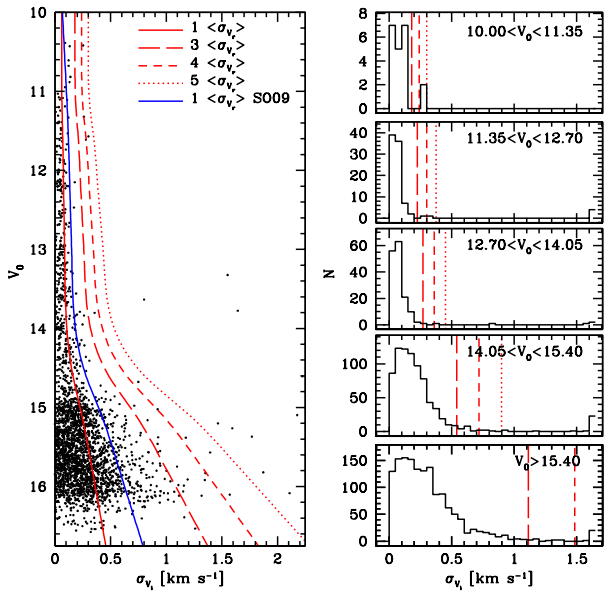}
\caption{The unbiased weighted radial velocity dispersion determined for 
   stars with multi-epoch measurements. 
   The estimated error on the radial velocity as a function of
   magnitude $\langle \sigma_{v_r}  \rangle (V)$ from our analysis (red
   thick line), is compared with the one obtained by
   \citet{Sommariva:2009cz} (blue line) using the same dataset. The histograms of the distribution for
   different ranges of magnitude are shown on the right panels,
   where the vertical lines follow the same
     convention defined in the left panel.
   Stars that satisfy the condition 
$\sigma_{v_r} > 3 \langle \sigma_{v_r} \rangle$ 
are considered binary candidates.}
\label{fig:RV_rms}
\end{figure}

The average scatter $\langle \sigma_{v_r} \rangle^{bin}$ 
for a magnitude bin is computed as the $68.27^{\rm th}$ percentile of 
$\sigma_{v_r}$ distribution.
This value is computed for several magnitude bins and
the resulting values are interpolated to determine the estimated error
in the radial velocity as a function of magnitude $\langle \sigma_{v_r}
\rangle(V)$ 
(the subscript is omitted for the sake of brevity in the next sections).
In the left-hand panel of Fig.~\ref{fig:RV_rms} 
we show the computed $\langle \sigma_{v_r}\rangle$ values for our M\,4 stars 
that have multi-epoch observations.
In addition, we show the comparable trend that was derived by 
S09 (red thick line in Fig.~10 of their paper,
blue line in Fig.~\ref{fig:RV_rms}). 
In the right-hand panel of this figure we show histograms of 
$\sigma_{v_r}$ in five magnitude bins.
To ensure a fair comparison, we did not
  include the data acquired in 2009 in this analysis (see Section \ref{sec:data-description}). 
At high \snr\ (\ie\ $V_0<13$) the error is dominated by the 
instrumental stability, the goodness of the SimCals correction and the 
precision of the RV determination technique, for a total uncertainty 
of $\simeq 100$~\ms\ ($<0.0018$ \AA, or $<0.02$ pixel). 
We regard this as the practical precision limit in
radial velocity that can be obtained from these GIRAFFE data.

For brighter (RGB) M\,4 stars the effect of the sky in RV determination is 
negligible even for observations obtained close to the full moon. 
For fainter stars with lower \snr\ data, the quality of the spectra 
becomes the main component in the precision of the RV determination.
The combination of an improved wavelength calibration and careful 
determinations of radial velocities result in an increased precision 
of about a factor of two compared to the 
S09 results
in the radial velocity determination of a single star.

Following S09, we consider stars with  
$\sigma_{v_r} > 3 \langle \sigma_{v_r} \rangle$ to be M\,4 binary
candidates. Using their same dataset, \ie, without data taken in 2009,
we find 22 candidates out of 454 targets with $V_0<14.9$, and 55 candidates 
out of 2068 targets with $V_0>14.9$, for a total of 77 binary candidates. 
A total of 20 more candidates are found compared to the previous analyses by
S09, using the same dataset and the same statistical 
approach. 
When the data from 2009 are included in the analysis 10
  additional binary candidates with  $V_0>14.9$ are found,  with the total number of
candidates growing to 87. Results are summarized in
Table~\ref{tab:rv_table}, while individual RVs are reported in table \ref{tab:rv_indiv_table}.

\begin{table*}
\center
\caption{Average radial velocities of the 2791 stars of the
  sample. Only a sample is given here, the full table will be
  available in the online version of the manuscript, together with a
  table comprising the binary candidates only.}
\label{tab:rv_table} 
\begin{tabular}{cccccccccccc}
\hline
$\alpha$~(J2000) & $\delta$~(J2000) & ID & $V$  & $(B-V)$ & $V_0$ & $(B-V)_0$ &
                                                                    No. Obs. &
                                                                               $v_r$ &
                                                                               $\sigma_{v_r}$
  & $\langle \sigma_{v_r} \rangle$ &    $\sigma_{v_r} / \langle \sigma_{v_r} \rangle$
  \\ 
&&&&&&&& \kms & \kms & \kms & \kms \\
\hline\hline
245.838250 & -26.513639 & 29622 & 16.329 & 0.970 & 14.974 & 0.615 &  2  & 70.553 & 0.115 & 0.221 &  0.521 \\
245.892417 & -26.504083 & 40918 & 14.505 & 1.139 & 13.146 & 0.771 &  2  & 70.963 & 0.023 & 0.088 &  0.266 \\
245.913167 & -26.483389 & 46022 & 15.016 & 1.103 & 13.649 & 0.742 &  3  & 63.619 & 0.084 & 0.094 &  0.891 \\
245.941333 & -26.560778 & 51878 & 12.935 & 1.282 & 11.570 & 0.930 & 10  & 67.314 & 0.266 & 0.069 &  3.834 \\
245.897083 & -26.537944 & 42121 & 12.944 & 1.245 & 11.566 & 0.873 & 10  & 79.513 & 6.666 & 0.069 & 96.145 \\
...&...&...&..&...&...&...&...&...&...&...&... \\
\hline
\end{tabular}
\end{table*}

\begin{table*}
\center
\caption{Individual radial velocities of the 2791 stars of the
  sample. Only a sample is given here, the full table will be
  available in the online version of the manuscript.}
\label{tab:rv_indiv_table} 
\begin{tabular}{ccccccccccc}
\hline
 ID  & MJD & j  & 
                                                                       GIRAFFE
                                                                       File
  & Fib & BERV & $v^{obs}_{r,j}$ & $v_{r,j}$  & \fwhm\ & \snr\ & $(\chi^2)^{-1}$ \\
\hline\hline
 34087  &  52819.117744  &  1 &  {\sc GIRAF.2003-06-29T02:49:33.104 } &  2  &  -13.951  &    83.002  &   69.051   &   9.971   &    7.25  &   105.998  \\
  34087  &  53982.999641  &  2  &  {\sc GIRAF.2006-09-04T23:59:28.940 } & 22  &  -29.533  &    99.591  &   70.058   &   9.226   &   12.08  &   171.798  \\
  30327  &  52819.117744  &  1  &  {\sc GIRAF.2003-06-29T02:49:33.104 } &  3  &  -13.957  &    86.523  &   72.566   &   9.048   &    8.61  &   134.988  \\
 30327  &  53947.065775  &  2  &  {\sc GIRAF.2006-07-31T01:34:42.930 } &  5  &  -25.373  &    98.630  &   73.257   &   8.734   &   16.38  &   185.676  \\
 30327  &  53981.986284  &  3  &  {\sc GIRAF.2006-09-03T23:40:14.963 } &  5  &  -29.544  &   102.366  &   72.822   &   9.420   &   12.66  &   161.065  \\
...&...&...&...&...&...&...&...&...&...&... \\
\hline
\end{tabular}
\end{table*}

Target stars have been carefully selected using multi-band photometry and proper
motions, drastically reducing the chance of contamination by other
stars. If present, contaminated spectra would have caused
anomalous values in the FWHM or depth of the fitting function, which
however have not been observed.

We did not attempt to determine the cluster binary fraction because
the limited number of available epochs requires detailed completeness
simulations, which are beyond the scope of our work, but it is likely
that the true fraction in the outskirts of the cluster is
substantially less than 0.1, in agreement with \cite{Milone:2012dr}
and \citet[Paper~III]{Nascimbeni:2014m4}. A future
  paper of the series will deal with the fraction and spatial
  distribution of binaries in M\~4 with a joint analysis of
  photometry, RV and proper motions.

\subsection{Mean radial velocity and velocity dispersion}\label{sec:mean-radial-velocity}

\begin{figure}
\includegraphics[width=\columnwidth]{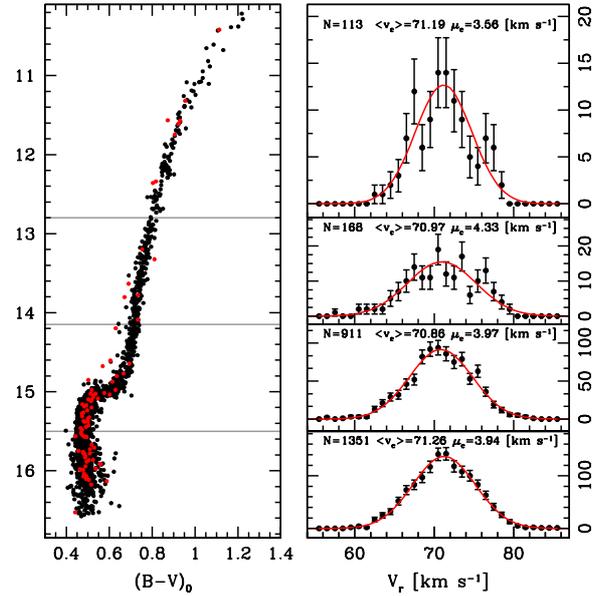}
\caption{The weighted radial velocity mean of the cluster, the 
  dispersion of the distribution and the number of stars
  for four magnitude bins, delimited with a grey line in the
  CMD. Candidate binary stars are denoted in red.
  There is no significant trend of the mean velocity of the cluster 
  with magnitude.}
\label{fig:fig015}
\end{figure}

The average velocity of M\,4 has been computed for four magnitude bins. 
Binary candidates are excluded from the sample, while we retain 
stars with only one RV measurement by assigning them 
$\sigma_{v_i} = \langle \sigma_{v_r} \rangle$ evaluated at their magnitude. 
Outliers in the CMD and obvious non-members (\ie\ those stars with
RV lower than 55~\kms\ or higher than 87~\kms) are excluded as well, 
leading to a total of 2543 stars out of the 2771 of the initial sample. 
The weighted radial velocity mean for the cluster, the associated error, 
the dispersion of the distribution, and the number of stars for each 
magnitude bin are reported in Table~\ref{tab:tab03} and 
displayed in Fig.~\ref{fig:fig015}. Errors on RV
  dispersions have been calculated as described in \S~\ref{sec:rad_vel_disp}
There is no significant trend
in the cluster mean radial velocity with magnitude.
 
\begin{table}
\center
\caption{Radial velocity means}
\label{tab:tab03} 
\begin{tabular}{lccccc}
\hline\hline
N & $V_0$ range & $\langle v_c \rangle$ & $\sigma_{\langle v_c \rangle}$ &
$\mu_{v_c}$ & $\sigma_{\mu_{v_c}}$ \\
  &         &                 \kms\ &                          \kms\ &
         \kms\ & \kms\\
\hline
113   & $10.10 \le V_0 < 12.8$ &71.19 & 0.33 & 3.56 & 0.24 \\
168   & $12.80 \le V_0 < 14.15$ &70.97 & 0.33 & 4.33 & 0.24 \\
911  & $14.15 \le V_0 < 15.50$ & 70.86 & 0.13 & 3.97 & 0.09 \\
1351  & $15.50 \le V_0 < 16.85$ &71.26 & 0.11 & 3.94 & 0.08 \\
totals: \\
2543 & $10.2 \le V_0 < 17.0$ & 71.08 & 0.08 &  3.97 & 0.05 \\
\hline
\end{tabular}
\end{table}

Our measured mean radial velocity for M\,4 is 
$\langle v_c \rangle = 71.08 \pm 0.08$~\kms\ with a 
cluster dispersion of $\mu_{v_c} = 3.97 \pm 0.05$~\kms.
This value is in agreement with previous studies. 
P95 reported a radial velocity mean of
$\langle v \rangle = 70.9\pm 0.6$~\kms\ using 200 giant stars. 
\cite{Cote:1996a1} derived a value of $70.3 \pm 0.7$~\kms\
from the analysis 33 turn-off dwarf stars. 
Our analysis marginally agrees with the one performed by
S09, which determined a mean radial velocity of
$70.27 \pm 0.19$~\kms\ using our same dataset. 
The main differences with that study, as described in previous sections, 
are in the wavelength calibration and in their use of a numerical mask. 
Unfortunately no radial velocity standards are present in our dataset, so for 
the moment it is not possible to assess the origin of this small offset. 

\begin{figure}
\includegraphics[width=\columnwidth]{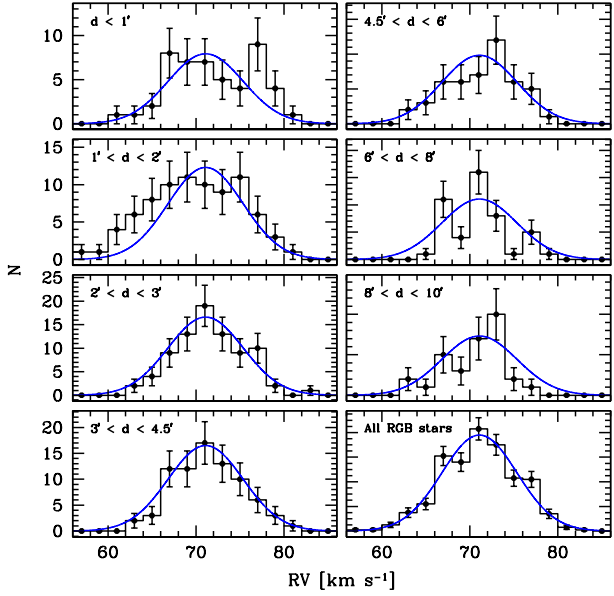}
\caption{RV distribution of RGB stars ($V_0<14.75$)
   for seven radial annuli (as defined in the label of each panel)  and for the whole
   dataset (last panel on the right side). The distribution obtained
   by including all the RGB stars, rescaled for the number of stars
   inside each annulus, is shown for comparison (blue curve).}
\label{fig:hist_RGB}
\end{figure}

Before discussing the RV dispersion as a function of the
  distance from cluster center, 
note from the two upper-right panels in Fig.~\ref{fig:fig015} that the RV distribution for RGB
and SGB stars deviates from a simple gaussian distribution. To verify that
these deviations do not affect the determination of the radial dispersion profile of the cluster, in Fig~\ref{fig:hist_RGB} we have plotted the RV distribution of RGB stars ($V_0<14.75$) for seven annuli of radial distance and for the full dataset (in the last panel on the right). The
gaussian fit obtained using the full RGB sample is shown for
comparison in all the panels, after properly rescaling for the total
number of stars in each sample.

While the deviation from the normal distribution of each sample
is not statistically significant, we have preferred to exclude the RGB
and SGB stars from the following analysis since in our judgement they
cannot provide a reliable picture of the kinematical status of the
cluster. It is also likely that evolved stars have different
kinematics with respect to MS stars as a consequence of mass
segregation, thus supporting our choice of analysing the MS sample
alone. The faintness of MS stars fortunately is not a limit in 
the determination of the kinematical properties of the cluster, thanks to
  the large size of the sample and our efforts in
improving the RV precision.

\subsection{Radial velocity dispersion}\label{sec:rad_vel_disp}

\begin{figure}
\includegraphics[width=\columnwidth]{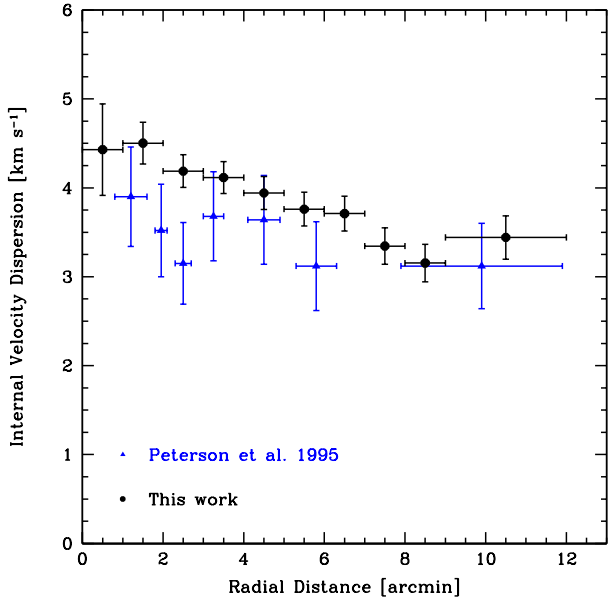}
\caption{The radial velocity dispersion is plotted as a function of 
  distance from the cluster center (black circles). 
Thanks to the larger sample and the improved RV
  precision, our data show that the velocity dispersion
  is steadily decreasing as a function of radius. This trend is
  suggested, but not statistically significant, by the data from
  \citet{Peterson:1995iy} (blue triangles)}
\label{fig:RVdispersion}
\end{figure}

The dependence of the radial velocity dispersion with the distance from 
the cluster center is analysed by dividing our sample into 10 radial,
circular annuli on the sky projected (tangent) plane and centered on
$(\alpha,\delta)_{\rm J2000.0}$$=$$(\rm 16^h23^m35^s\!\!.22,$$\rm
-26^{\circ}31^{\prime}32^{\prime\prime}\!\!.7)$ \citep{Goldsbury:2010cc}. The
weighted velocity dispersion in each bin
is then calculated, as 
displayed in Fig.~\ref{fig:RVdispersion}.

To take into account the broadening due to the observational uncertainty, 
for each bin the $68.27^{\rm th}$ percentile of the error distribution has been 
computed and the resulting value ($\simeq 0.3$~\kms) 
quadratically subtracted from the observed cluster dispersion
$\mu_{v_c}$ in each bin. 
 The error in the mean has been quadratically subtracted as well.
We use the symbol $\mu_{\rm int}$ to distinguish the resulting value from 
the observed cluster dispersion. 
The vertical error for each bin has been determined following \cite{Peterson:1986cp} 
for a consistent comparison with P95 measurements. 
The square in the uncertainty in $\mu_{\rm int}^2$ has been computed as the 
sum of sampling error $\sigma_N^2=2/N \mu_{\rm int}^4 $, where $N$ the number 
of stars in each bin, plus the error due to the uncertainty
in a single measurements $\sigma_{\mu}^2 = 4/(N-1)  \langle
\sigma_{v_r} ^2  \rangle   \sigma_{v_c}^2$. 
Following the propagation of uncertainty, the error in $\mu_{\rm int}$ is 
given by dividing the sum resulting from the previous step by $2
\mu_{\rm int}$. Results are summarized in Table~\ref{tab:rvdisp}.

\begin{table}
\center
\caption{Radial velocity dispersion as function of radial distance} 
\label{tab:rvdisp} 
\begin{tabular}{cccccc} 
\hline\hline
$r $ & $\sigma_r$& $\mu_{v_c}$ &  $\mu_{\rm int}$
& $\sigma_{\mu_{\rm int}}$ & No. Stars \\
arcmin & arcmin & \kms\ & \kms\ & \kms\ &\\
\hline
0.5 &0.3  &4.50 &4.43 &0.52 & 40   \\
1.5 &0.3  &4.53 &4.50 &0.24 & 189  \\
2.5 &0.4  &4.21 &4.19 &0.19 & 268  \\
3.5 &0.3  &4.14 &4.12 &0.18 & 271  \\
4.5 &0.3  &3.97 &3.94 &0.19 & 233  \\
5.5 &0.3  &3.79 &3.76 &0.19 & 205  \\
6.5 &0.3  &3.74 &3.71 &0.20 & 189  \\
7.5 &0.3  &3.36 &3.35 &0.20 & 143  \\
8.5 &0.4  &3.19 &3.16 &0.21 & 119  \\
10.5&1.2  &3.48 &3.43 &0.24 & 106  \\
\hline
\end{tabular}
\end{table}

Thanks to the larger sample and the improved RV
  precision, our data show a steadily decreasing trend of the radial
  velocity dispersion as a function of radius. Our data are compatible
  with the one of P95, although in their case the
  trend is not statistically significant.
This trend is still present with the same amplitude under different 
assumed statistical restrictions.
These include: \textit{(a)} considering only stars with at least three 
individual radial velocities; \textit{(b)} applying 
a different $\sigma$-clipping on RV errors, \ie\ $\sigma_{v_r} \le 3 \langle
\sigma_{v_r} \rangle$ instead of  $\sigma_{v_r} \le 5 \langle \sigma_{v_r}
\rangle$; \textit{(c)} selecting the stars in magnitude intervals. 
The radial distribution in P95 was determined using
182 members with a radial velocity precision of 
$\simeq 1$~\kms\ for individual stars, while our sample 
comprises a larger number of stars (1763 MS members) and higher precision 
for a single star velocity determination (a few hundred \ms).

\subsection{Rotation}\label{sec:rotation}
\begin{figure}
\includegraphics[width=\columnwidth]{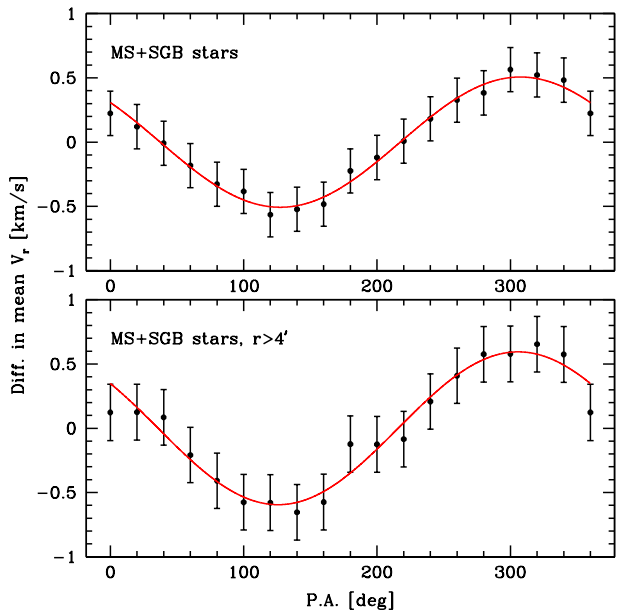}
\caption{The difference between the mean velocities of the two halves
  of the cluster when divided by a line with varying PA, as a function
  of the PA itself. The best-fitting sine curve is shown. The analysis
  is performed on a sample including MS and
SGB stars only (upper panel), and the same sample with only stars
farther than 4$^{\prime}$ from the 
  cluster center (lower panel).}
\label{fig:rot_curve}
\end{figure}

Cluster rotation has been determined by following the same approach of
\cite{Cote:1995dn}. Briefly, the cluster is halved by the position
angle ${\rm PA}$,  and the difference between the mean radial velocity of the two sides is computed, the error associated to each point being the sum in quadrature of the error on the mean of each side. 
The rotation curve as a function of the ${\rm PA}$ is then obtained by
repeating the measurement for different values of the ${\rm PA}$. We adopted
the same convention of Bellazzini et~al. (2012; hereafter B12)\nocite{Bellazzini:2012bg}, \ie\ the ${\rm PA}$ is
increasing anti-clockwise from north (${\rm PA}=0^{\circ}$) to east
(${\rm PA}=90^{\circ}$) in the plane of the sky, and the sine-function to
fit the curve as $\Delta V_r = A_{\rm rot} \sin({\rm PA}+ 270
-{\rm PA}_0$), where ${\rm PA}_0$
corresponds to the ${\rm PA}$ of the rotation axis and
$A_{\rm rot}$ is the rotation
amplitude projected on the plane of the sky and averaged over the
range of radius of the sample. 
To better investigate the properties of the cluster, we have performed
the analysis twice, by including all MS and SGB stars (which should
share the same dynamical properties), and then by considering only
only stars farther than $4^{\prime}$ from the cluster center, as
previously done
by P95. 
Binary candidates identified in Sec. \ref{sec:results-discussion} have been excluded from the analysis.
The best-fit values are $A_{\rm rot} = 0.51 \pm 0.06$ \kms\ and ${\rm PA}_0 =
127 \pm 7 ^{\circ}$  for the whole sample, and $A_{\rm rot} = 0.60 \pm
0.07$ \kms\ and ${\rm PA}_0 = 125 \pm 7 ^{\circ}$ for the
distance-selected one, thus being perfectly consistent. 
(Fig. \ref{fig:rot_curve}). These values are obtained using a
Monte Carlo method. Each data point is perturbed of a random
value extracted from a normal distribution with the associated error
as standard rotation, and the rotation curve is fitted. This procedure is then
iterated 10.000 times to obtain a distribution of the two parameters
and an estimate for their errors.  

\begin{figure}
\includegraphics[width=\columnwidth]{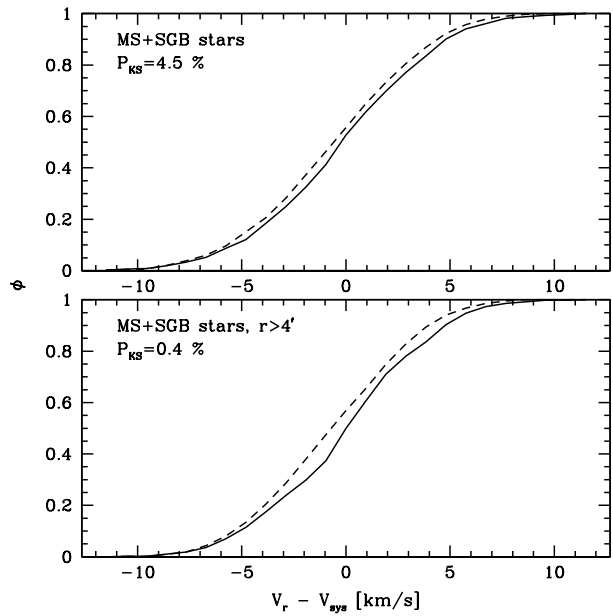}
\caption{Comparison of the cumulative $v_r$ distribution of stars on
  the left (dashed lines) and on the right (continuous lines) side of
  the rotation axis of the cluster. The cumulative distributions along
  with the probability of null-hypothesis of the Kolmogorov-Smirnov
  test $P_{\rm KS}$ is shown for the whole MS$+$SGB sample (upper
  plot), and for MS$+$SGB stars farther than $4\arcsec$ from the
  cluster center. This result confirms the moderate rotation already
  observed in previous works.}
\label{fig:rot_cumul}
\end{figure}

A common test to assess the statistical significance of
the rotation signal is to perform a two-side Kolmogorov-Smirnov test
on the cumulative distribution of RVs of stars that lie on opposite
sides of the rotation axis (B12, \citealt{Lardo:2015ll}). The presence
of rotation induces a shift between the two distribution, see
Fig. \ref{fig:rot_cumul}, hence the KS 
test quantify the probability to obtain the observed distribution
under the null hypothesis that no rotation is present. We obtain 
$P_{\rm KS}=4.5 \% $ and $P_{\rm KS}=0.4 \% $ for the full sample and the
distance-selected one respectively. The lower probability in the
second case is expected, since we have removed from the sample the
central region of the cluster, where the projected rotational velocity
becomes negligible. 
For the RGB stars alone we obtain $P_{\rm KS} > 20 \% $, so we have
not included the analysis in the text.
Our results are consistent with the previous claim
of moderate rotation for this cluster of P95. We found a rotation
amplitude several times smaller than the value of $A_{\rm rot} = 1.8 \pm
0.2 $ \kms of \cite{Lane:2010cm}, although they quote half of this
value as the rotation velocity.

\section{Conclusions}\label{sec:conclusions}

We have presented a new approach for the determination of accurate
wavelength dispersion solutions for not-stabilized instruments. 
We have developed an algorithm for improved corrections for radial 
velocity shifts using simultaneous calibration fibres. 
Several precautions have to be taken when dealing with low \snr\
spectra covering a short wavelength range and strong absorption
features, in particular during sky spectrum subtraction and continuum
normalization. 
Together with a careful choice of the radial velocity technique that better 
suites the characteristics of our spectra, we have been able to significantly 
increase the precision of the radial velocities with respect to existing tools. 
Algorithms have been optimized to require only a minimal user
interaction. 

The code we have developed is lacking the necessary documentation to
be user-friendly at present. While we do not exclude a general release in the next future, we have decided to not
make it public yet. Instead, we provide a very detailed description of each step of our methodology
with the hope that large consortium dedicated to data reduction, such
as the Gaia-ESO or ESO itself, can implement and eventually improve it
for other GIRAFFE settings as well other instruments.

Our new radial velocity procedures have been tested on a large dataset for 
globular cluster M\,4 with extant state-of-the-art radial velocity 
measurements in the literature, in order to assess the improvement 
in velocity precision. 
A total of 7250 individual spectra for 2771 stars, gathered with the
GIRAFFE spectrograph at VLT during a multi-epoch campaign, have been
analysed, for a total of 2543 stars after removing non-members and
candidate binaries.
The derived cluster radial velocity is 
$\langle v \rangle = 71.08 \pm 0.08$~\kms, in 
agreement with previous measurements in
literature but with an offset of $\Delta v =0.81$~\kms\
with the previous value derived using the same dataset (probably due
to the different radial velocity determination technique). 
We found a total of 87 binary
 candidates, 22 giants and 65 sub-giants and dwarfs, which is a  significative
 increase of the number found by 
previous analyses.

From our data we have measured a rotational amplitude
  of $\simeq 0.5$\kms, providing a statistical confirmation of the
  negligible rotation of this cluster.
An average radial velocity dispersion of $4.5$~\kms\ within
2$\arcmin$ from the center of cluster and steadily decreasing 
outward has been found, in contrast with the nearly constant value of  
$3.5$~\kms\ from P95. Our new
  determination is going in the same direction of recent simulations
  from \cite{Heggie:2014m4}, where a N-body simulation of the cluster results in a RV dispersion steadily, and it will further improve the dynamical modeling of M4. 
Ultimately, the combination of these data with the unprecedented
precision of the proper-motions that will be delivered by the \textit{HST}
Large Program described in Paper~I and following will provide a
deeper insight into the three-dimensional dynamics of the cluster 
and a new, more accurate determination of its geometrical distance,
among other things.

\section*{Acknowledgments}
LM acknowledges the financial support provided by \textit{Fondazione
Ing. Aldo Gini} and from the European Union Seventh Framework Programme (FP7/2007-2013) under Grant agreement number 313014 (ETAEARTH).
LRB, GP, VN and LM acknowledge PRIN-INAF 2012
funding under the project entitled \textit{The M 4 Core Project with
  Hubble Space Telescope}. Partial support for this work has been provided by the US National Science
Foundation under grant AST-1211585.
We wish to thank the anonymous referee for the detailed reading and the useful comments which helped to improve significantly  the paper.

\bibliographystyle{mn2e} 
\bibliography{Bibliography}

\begin{thebibliography}{60}
\expandafter\ifx\csname natexlab\endcsname\relax\def\natexlab#1{#1}\fi

\bibitem[{Anderson {et~al}\mbox{.}(2006)Anderson, Bedin, Piotto, Yadav, \&
  Bellini}]{Anderson:2006a}
Anderson J., Bedin R.~R., Piotto G., Yadav R.~S., Bellini A., 2006, A{\&}A,
  454, 1029

\bibitem[{Anderson \& King(2000)}]{Anderson:2000bt}
Anderson J., King I.~R., 2000, PASP, 112, 1360

\bibitem[{Baranne {et~al}\mbox{.}(1996)Baranne, Queloz, Mayor, Adrianzyk,
  Knispel, Kohler, Lacroix, Meunier, Rimbaud, \& Vin}]{Baranne:1996tb}
Baranne A. {et~al.}, 1996, Astron Astrophys Sup, 119, 373

\bibitem[{{Baumgardt} {et~al}\mbox{.}(2009){Baumgardt}, {C{\^o}t{\'e}},
  {Hilker}, {Rejkuba}, {Mieske}, {Djorgovski}, \& {Stetson}}]{Baumgardt:2009cd}
{Baumgardt} H., {C{\^o}t{\'e}} P., {Hilker} M., {Rejkuba} M., {Mieske} S.,
  {Djorgovski} S.~G., {Stetson} P., 2009, \mnras, 396, 2051

\bibitem[{{Bedin} {et~al}\mbox{.}(2013){Bedin}, {Anderson}, {Heggie}, {Piotto},
  {Milone}, {Giersz}, {Nascimbeni}, {Bellini}, {Rich}, {van den Berg},
  {Pooley}, {Brogaard}, {Ortolani}, {Malavolta}, {Ubeda}, \&
  {Marino}}]{Bedin:2013cd}
{Bedin} L.~R. {et~al.}, 2013, Astronomische Nachrichten, 334, 1062

\bibitem[{Bellazzini {et~al}\mbox{.}(2012)Bellazzini, Bragaglia, Carretta,
  Gratton, Lucatello, Catanzaro, \& Leone}]{Bellazzini:2012bg}
Bellazzini M., Bragaglia A., Carretta E., Gratton R., Lucatello S., Catanzaro
  G., Leone F., 2012, A{\&}A, 538, 18

\bibitem[{Blecha {et~al}\mbox{.}(2000)Blecha, Cayatte, North, Royer, \&
  Simond}]{Blecha:2000ud}
Blecha A., Cayatte V., North P., Royer F., Simond G., 2000, Proc. SPIE Vol.
  4008, 4008, 467

\bibitem[{{Bristow} {et~al}\mbox{.}(2010){Bristow}, {Vernet}, {Kerber},
  {Moehler}, \& {Modigliani}}]{Bristow2010:sp}
{Bristow} P., {Vernet} J., {Kerber} F., {Moehler} S., {Modigliani} A., 2010, in
  Society of Photo-Optical Instrumentation Engineers (SPIE) Conference Series,
  Vol. 7735, Society of Photo-Optical Instrumentation Engineers (SPIE)
  Conference Series, p.~7

\bibitem[{Brown \& Wallerstein(1992)}]{Brown:1992a}
Brown J.~A., Wallerstein G., 1992, AJ, 104, 1818

\bibitem[{Casagrande {et~al}\mbox{.}(2010)Casagrande, Ram{\'\i}rez,
  Mel{\'e}ndez, Bessell, \& Asplund}]{Casagrande:2010a}
Casagrande L., Ram{\'\i}rez I., Mel{\'e}ndez J., Bessell M., Asplund M., 2010,
  A{\&}A, 512, 54

\bibitem[{Castelli \& Kurucz(2004)}]{Castelli:2004ti}
Castelli F., Kurucz R.~L., 2004, arXiv, astro-ph

\bibitem[{C{\^o}t{\'e} \& Fischer(1996)}]{Cote:1996a1}
C{\^o}t{\'e} P., Fischer P., 1996, Astronomical Journal v.112, 112, 565

\bibitem[{C{\^o}t{\'e} {et~al}\mbox{.}(1995)C{\^o}t{\'e}, Welch, Fischer, \&
  Gebhardt}]{Cote:1995dn}
C{\^o}t{\'e} P., Welch D.~L., Fischer P., Gebhardt K., 1995, Astrophysical
  Journal v.454, 454, 788

\bibitem[{D'Antona {et~al}\mbox{.}(2009)D'Antona, Stetson, Ventura, Milone,
  Piotto, \& Caloi}]{DAntona:2009kg}
D'Antona F., Stetson P.~B., Ventura P., Milone A.~P., Piotto G., Caloi V.,
  2009, Monthly Notices of the Royal Astronomical Society: Letters, L319

\bibitem[{de~Boor(1977)}]{Boor:1977:PCB}
de~Boor C., 1977, Journal of Numerical Analysis, 14, 441

\bibitem[{{Dinescu} {et~al}\mbox{.}(1999){Dinescu}, {van Altena}, {Girard}, \&
  {L{\'o}pez}}]{Dinescu:1999cd}
{Dinescu} D.~I., {van Altena} W.~F., {Girard} T.~M., {L{\'o}pez} C.~E., 1999,
  \aj, 117, 277

\bibitem[{{D'Orazi} {et~al}\mbox{.}(2013){D'Orazi}, {Campbell}, {Lugaro},
  {Lattanzio}, {Pignatari}, \& {Carretta}}]{Dorazi:2013a}
{D'Orazi} V., {Campbell} S.~W., {Lugaro} M., {Lattanzio} J.~C., {Pignatari} M.,
  {Carretta} E., 2013, \mnras, 433, 366

\bibitem[{Enard(1982)}]{Enard:1982wf}
Enard D., 1982, Society of Photo-Optical Instrumentation Engineers (SPIE)
  Conference Series, 331, 232

\bibitem[{{Feldmeier} {et~al}\mbox{.}(2013){Feldmeier}, {L{\"u}tzgendorf},
  {Neumayer}, {Kissler-Patig}, {Gebhardt}, {Baumgardt}, {Noyola}, {de Zeeuw},
  \& {Jalali}}]{Feldmeier:2013cd}
{Feldmeier} A. {et~al.}, 2013, \aap, 554, A63

\bibitem[{{Goldsbury} {et~al}\mbox{.}(2010){Goldsbury}, {Richer}, {Anderson},
  {Dotter}, {Sarajedini}, \& {Woodley}}]{Goldsbury:2010cc}
{Goldsbury} R., {Richer} H.~B., {Anderson} J., {Dotter} A., {Sarajedini} A.,
  {Woodley} K., 2010, \aj, 140, 1830

\bibitem[{{Heggie}(2014)}]{Heggie:2014m4}
{Heggie} D.~C., 2014, \mnras, 445, 3435

\bibitem[{Heggie \& Giersz(2008)}]{Heggie:2008c5}
Heggie D.~C., Giersz M., 2008, Monthly Notice of the Royal Astronomical
  Society, 389, 1858

\bibitem[{Hendricks {et~al}\mbox{.}(2012)Hendricks, Stetson, Vandenberg, \&
  Dall'Ora}]{Hendrick:2012cw}
Hendricks B., Stetson P.~B., Vandenberg D.~A., Dall'Ora M., 2012, AJ, 144, 25

\bibitem[{Horne(1986)}]{Horne:1986bg}
Horne K., 1986, PASP, 98, 609

\bibitem[{{Ibata} {et~al}\mbox{.}(2011){Ibata}, {Sollima}, {Nipoti},
  {Bellazzini}, {Chapman}, \& {Dalessandro}}]{Ibata:2011cd}
{Ibata} R., {Sollima} A., {Nipoti} C., {Bellazzini} M., {Chapman} S.~C.,
  {Dalessandro} E., 2011, \apj, 738, 186

\bibitem[{Ivans {et~al}\mbox{.}(1999)Ivans, Sneden, Kraft, Suntzeff, Smith,
  Langer, \& Fulbright}]{Ivans:1999hf}
Ivans I.~I., Sneden C., Kraft R.~P., Suntzeff N.~B., Smith V.~V., Langer G.~E.,
  Fulbright J.~P., 1999, AJ, 118, 1273

\bibitem[{{Kaluzny} {et~al}\mbox{.}(2013){Kaluzny}, {Thompson}, {Rozyczka},
  {Dotter}, {Krzeminski}, {Pych}, {Rucinski}, {Burley}, \&
  {Shectman}}]{Kaluzny:2013cd}
{Kaluzny} J. {et~al.}, 2013, \aj, 145, 43

\bibitem[{Kirby, Guhathakurta \& Sneden(2008)Kirby, Guhathakurta, \&
  Sneden}]{Kirby:2008cr}
Kirby E., Guhathakurta P., Sneden C., 2008, ApJ, 682, 1217

\bibitem[{Kurtz \& Mink(1998)}]{Kurtz:1998aa}
Kurtz M.~J., Mink D.~J., 1998, PASP, 110, 934

\bibitem[{Kurucz(1992)}]{Kurucz:1992ab}
Kurucz R.~L., 1992, in The Stellar Populations of Galaxies: Proceedings of the
  149th Symposium of the International Astronomical Union, p. 225

\bibitem[{{Kurucz} {et~al}\mbox{.}(1984){Kurucz}, {Furenlid}, {Brault}, \&
  {Testerman}}]{Kurucz:1984ab}
{Kurucz} R.~L., {Furenlid} I., {Brault} J., {Testerman} L., 1984, {Solar flux
  atlas from 296 to 1300 nm}. National Solar Observatory

\bibitem[{Lane {et~al}\mbox{.}(2010)Lane, Kiss, Lewis, Ibata, Siebert, Bedding,
  Sz{\'e}kely, Balog, \& Szab{\'o}}]{Lane:2010cm}
Lane R.~R. {et~al.}, 2010, Monthly Notice of the Royal Astronomical Society,
  406, 2732

\bibitem[{Lardo {et~al}\mbox{.}(2015)Lardo, Pancino, Bellazzini, Bragaglia,
  Donati, Gilmore, Randich, Feltzing, Jeffries, Vallenari, Alfaro,
  Allende~Prieto, Flaccomio, Koposov, Recio-Blanco, Bergemann, Carraro,
  Costado, Damiani, Hourihane, Jofr{\'e}, de~Laverny, Marconi, Masseron,
  Morbidelli, Sacco, \& Worley}]{Lardo:2015ll}
Lardo C. {et~al.}, 2015, A{\&}A, 573, A115

\bibitem[{{Libralato} {et~al}\mbox{.}(2014){Libralato}, {Bellini}, {Bedin},
  {Piotto}, {Platais}, {Kissler-Patig}, \& {Milone}}]{Libralato:2014m4}
{Libralato} M., {Bellini} A., {Bedin} L.~R., {Piotto} G., {Platais} I.,
  {Kissler-Patig} M., {Milone} A.~P., 2014, \aap, 563, A80

\bibitem[{Lovis \& Pepe(2007)}]{Lovis:2007ac}
Lovis C., Pepe F., 2007, A{\&}A, 468, 1115

\bibitem[{{L{\"u}tzgendorf} {et~al}\mbox{.}(2013){L{\"u}tzgendorf},
  {Kissler-Patig}, {Gebhardt}, {Baumgardt}, {Noyola}, {de Zeeuw}, {Neumayer},
  {Jalali}, \& {Feldmeier}}]{Lutzgendorf:2013cd}
{L{\"u}tzgendorf} N. {et~al.}, 2013, \aap, 552, A49

\bibitem[{Malavolta {et~al}\mbox{.}(2014)Malavolta, Sneden, Piotto, Milone,
  Bedin, \& Nascimbeni}]{Malavolta:2014hu}
Malavolta L., Sneden C., Piotto G., Milone A.~P., Bedin R.~R., Nascimbeni V.,
  2014, AJ, 147, 25

\bibitem[{Marino {et~al}\mbox{.}(2008)Marino, Villanova, Piotto, Milone,
  Momany, Bedin, \& Medling}]{Marino:2008du}
Marino A.~F., Villanova S., Piotto G., Milone A.~P., Momany Y., Bedin R.~R.,
  Medling A.~M., 2008, A{\&}A, 490, 625

\bibitem[{{Milone} {et~al}\mbox{.}(2014){Milone}, {Marino}, {Bedin}, {Piotto},
  {Cassisi}, {Dieball}, {Anderson}, {Jerjen}, {Asplund}, {Bellini}, {Brogaard},
  {Dotter}, {Giersz}, {Heggie}, {Knigge}, {Rich}, {van den Berg}, \&
  {Buonanno}}]{Milone:2014ab}
{Milone} A.~P. {et~al.}, 2014, \mnras, 439, 1588

\bibitem[{Milone {et~al}\mbox{.}(2012)Milone, Piotto, Bedin, Aparicio,
  Anderson, Sarajedini, Marino, Moretti, Davies, Chaboyer, Dotter, Hempel,
  Mar{\'\i}n-Franch, Majewski, Paust, Reid, Rosenberg, \&
  Siegel}]{Milone:2012dr}
Milone A.~P. {et~al.}, 2012, A{\&}A, 540, 16

\bibitem[{Milone {et~al}\mbox{.}(2006)Milone, Villanova, Bedin, Piotto,
  Carraro, Anderson, King, \& Zaggia}]{Milone:2006cc}
Milone A.~P., Villanova S., Bedin R.~R., Piotto G., Carraro G., Anderson J.,
  King I.~R., Zaggia S., 2006, A{\&}A, 456, 517

\bibitem[{{Monelli} {et~al}\mbox{.}(2013){Monelli}, {Milone}, {Stetson},
  {Marino}, {Cassisi}, {del Pino Molina}, {Salaris}, {Aparicio}, {Asplund},
  {Grundahl}, {Piotto}, {Weiss}, {Carrera}, {Cebri{\'a}n}, {Murabito},
  {Pietrinferni}, \& {Sbordone}}]{Monelli:2013ab}
{Monelli} M. {et~al.}, 2013, \mnras, 431, 2126

\bibitem[{Murphy {et~al}\mbox{.}(2007)Murphy, Tzanavaris, Webb, \&
  Lovis}]{Murphy:2007hn}
Murphy M.~T., Tzanavaris P., Webb J.~K., Lovis C., 2007, Monthly Notice of the
  Royal Astronomical Society, 378, 221

\bibitem[{{Nardiello} {et~al}\mbox{.}(2015){Nardiello}, {Milone}, {Piotto},
  {Marino}, {Bellini}, \& {Cassisi}}]{Nardiello2015:m4}
{Nardiello} D., {Milone} A.~P., {Piotto} G., {Marino} A.~F., {Bellini} A.,
  {Cassisi} S., 2015, \aap, 573, A70

\bibitem[{{Nascimbeni} {et~al}\mbox{.}(2014){Nascimbeni}, {Bedin}, {Heggie},
  {van den Berg}, {Giersz}, {Piotto}, {Brogaard}, {Bellini}, {Milone}, {Rich},
  {Pooley}, {Anderson}, {Ubeda}, {Ortolani}, {Malavolta}, {Cunial}, \&
  {Pietrinferni}}]{Nascimbeni:2014m4}
{Nascimbeni} V. {et~al.}, 2014, \mnras, 442, 2381

\bibitem[{Pasquini {et~al}\mbox{.}(2000)Pasquini, Avila, Allaert, Ballester,
  Biereichel, Buzzoni, Cavadore, Dekker, Delabre, Ferraro, Hill, Kaufer,
  Kotzlowski, Lizon, Longinotti, Moureau, Palsa, \& Zaggia}]{Pasquini:2000tk}
Pasquini L. {et~al.}, 2000, Society of Photo-Optical Instrumentation Engineers
  (SPIE) Conference Series, 4008, 129

\bibitem[{Pepe {et~al}\mbox{.}(2002)Pepe, Mayor, Galland, Naef, Queloz, Santos,
  Udry, \& Burnet}]{Pepe:2002ab}
Pepe F., Mayor M., Galland F., Naef D., Queloz D., Santos N.~C., Udry S.,
  Burnet M., 2002, A{\&}A, 388, 632

\bibitem[{Pepe {et~al}\mbox{.}(2004)Pepe, Mayor, Queloz, Benz, Bonfils, Bouchy,
  Lo~Curto, Lovis, Megevand, Moutou, Naef, Rupprecht, Santos, Sivan, Sosnowska,
  \& Udry}]{Pepe:2004hi}
Pepe F. {et~al.}, 2004, A{\&}A, 423, 385

\bibitem[{Peterson \& Latham(1986)}]{Peterson:1986cp}
Peterson R.~C., Latham D.~W., 1986, Astrophysical Journal, 305, 645

\bibitem[{Peterson, Rees \& Cudworth(1995)Peterson, Rees, \&
  Cudworth}]{Peterson:1995iy}
Peterson R.~C., Rees R.~F., Cudworth K.~M., 1995, ApJ, 443, 124

\bibitem[{{Piotto} {et~al}\mbox{.}(2015){Piotto}, {Milone}, {Bedin},
  {Anderson}, {King}, {Marino}, {Nardiello}, {Aparicio}, {Barbuy}, {Bellini},
  {Brown}, {Cassisi}, {Cool}, {Cunial}, {Dalessandro}, {D'Antona}, {Ferraro},
  {Hidalgo}, {Lanzoni}, {Monelli}, {Ortolani}, {Renzini}, {Salaris},
  {Sarajedini}, {van der Marel}, {Vesperini}, \& {Zoccali}}]{Piotto:2015uv}
{Piotto} G. {et~al.}, 2015, \aj, 149, 91

\bibitem[{Ram{\'\i}rez \& Mel{\'e}ndez(2005)}]{Ramirez:2005bz}
Ram{\'\i}rez I., Mel{\'e}ndez J., 2005, ApJ, 626, 465

\bibitem[{Royer {et~al}\mbox{.}(2002)Royer, Blecha, North, Simond, Baratchart,
  Cayatte, Chemin, \& Palsa}]{Royer:2002hf}
Royer F., Blecha A., North P., Simond G., Baratchart S., Cayatte V., Chemin L.,
  Palsa R., 2002, Astronomical Data Analysis II. Edited by Starck, 4847, 184

\bibitem[{Smith \& Briley(2005)}]{Smith:2005a}
Smith G.~H., Briley M.~M., 2005, PASP, 117, 895

\bibitem[{Sneden(1973)}]{Sneden:1973el}
Sneden C., 1973, Astrophysical Journal, 184, 839

\bibitem[{Sommariva {et~al}\mbox{.}(2009)Sommariva, Piotto, Rejkuba, Bedin,
  Heggie, Mathieu, \& Villanova}]{Sommariva:2009cz}
Sommariva V., Piotto G., Rejkuba M., Bedin R.~R., Heggie D.~C., Mathieu R.~D.,
  Villanova S., 2009, A{\&}A, 493, 947

\bibitem[{Sousa {et~al}\mbox{.}(2007)Sousa, Santos, Israelian, Mayor, \&
  Monteiro}]{Sousa:2007gn}
Sousa S.~G., Santos N.~C., Israelian G., Mayor M., Monteiro M. J. P. F.~G.,
  2007, A{\&}A, 469, 783

\bibitem[{Stumpff(1980)}]{Stumpff:1980ab}
Stumpff P., 1980, Astron Astrophys Sup, 41, 1

\bibitem[{Tonry \& Davis(1979)}]{Tonry:1979aa}
Tonry J., Davis M., 1979, Astronomical Journal, 84, 1511

\bibitem[{Yong {et~al}\mbox{.}(2008)Yong, Karakas, Lambert, Chieffi, \&
  Limongi}]{Yong:2008a}
Yong D., Karakas A.~I., Lambert D.~L., Chieffi A., Limongi M., 2008, ApJ, 689,
  1031

\end{thebibliography}
\bsp

\label{lastpage}
\end{document}